\documentclass
[preprint,tightenlines,prc,bibnotes,showpacs,noshowkeys,nofootinbib]{revtex4}%
\usepackage{amsfonts}
\usepackage{amsmath}
\usepackage{amssymb}
\usepackage{graphicx}%
\begin{document}
\preprint{JLAB-THY-05-01}
\preprint{NT@UW-04-030}
\title[lfcbm lattice form factor]{Comparison of Nucleon Form Factors from Lattice QCD Against the Light Front
Cloudy Bag Model and Extrapolation to the Physical Mass Regime \ }
\author{Hrayr H. Matevosyan}
\affiliation{Louisiana State University, Department of Physics \& Astronomy, 202 Nicholson
Hall, Tower Dr., LA 70803, USA, / Thomas Jefferson National Accelerator
Facility,12000 Jefferson Ave., Newport News, VA 23606, USA}
\author{Gerald A. Miller}
\affiliation{University of Washington, Department of Physics, Box 351560, Seattle, WA
98195-1560, USA}
\author{Anthony W. Thomas}
\affiliation{Thomas Jefferson National Accelerator Facility, 12000 Jefferson Ave., Newport
News, VA 23606, USA }
\keywords{Nucleon electromagnetic form factors, lfcbm, lattice QCD}
\pacs{13.40.Gp; 11.15.Ha; 12.38.Gc}

\begin{abstract}
We explore the possibility of extrapolating state of the art lattice QCD
calculations of nucleon form factors to the physical regime. We find that the
lattice results can be reproduced using the Light Front Cloudy Bag Model by
letting its parameters be analytic functions of the quark mass. We then use
the model to extend the lattice calculations to large values of $Q^{2}$ of
interest to current and planned experiments. These functions are also used to
define extrapolations to the physical value of the pion mass, thereby allowing
us to study how the predicted zero in $G_{E}(Q^{2})/G_{M}(Q^{2})$ varies as a
function of quark mass.

\end{abstract}
\volumeyear{year}
\volumenumber{number}
\issuenumber{number}
\eid{identifier}
\startpage{1}
\endpage{31}
\maketitle

\section{Introduction}

The electromagnetic form factors of the nucleon are an invaluable source of
information on its structure~\cite{Thomas:2001kw}. For example, observing
their fall as $Q^{2}$ increases from zero revealed the finite extent of the
nucleon, and measuring the Sachs electric form factor of the neutron,
$G_{E}^{n}$~\cite{Reichelt:2003iw,Glazier:2004ny}, that it has a positive core
surrounded by a long-range, negative
tail~\cite{Friedrich:2003iz,Thomas:1982kv,cbm}. In the last few years
particular interest has focused on the ratio of the electric and magnetic form
factors of the proton, $G_{E}/G_{M}$, where recoil polarization
data~\cite{Jones:1999rz,Gayou:2001qd} have revealed a dramatic decrease with
$Q^{2}$ -- in contrast with earlier work based on the Rosenbluth separation.
These data have allowed one of us to deduce a {}fascinating spin dependence of
the shape of the nucleon~\cite{Miller:2003sa}.

While the behavior of $G_{E}/G_{M}$ with $Q^{2}$ was anticipated in some
models (e.g. see Ref.~\cite{Frank:1995pv}, \cite{Chung:1991st}), there is no
consensus as to which explanation best represents how QCD works. Direct
guidance from QCD itself would be most valuable and for that purpose lattice
QCD represents the one and only technique by which one can obtain
non-perturbative solutions to QCD.

The QCDSF Collaboration recently presented lattice QCD simulations for the
{}form factors of the nucleon over a wide range of values of momentum
transfer~\cite{QCDSF Collaboration}. While these were based on the quenched
approximation, with an unsophisticated action, several lattice spacings were
chosen with the smallest being around $0.05%
\operatorname{fm}%
$ ($\beta=6.4$) and at present these are the state of the art. The quark
masses used in the simulations correspond to pion masses in the range
$(0.6,1.2)%
\operatorname{GeV}%
$. Therefore one needs to parametrize the form factors as a function of pion
mass and extrapolate to the physical value before comparing these lattice
results with the experimental data.

At $Q^{2}=0$ there have been a number of studies of the chiral extrapolation
of baryon magnetic moments
~\cite{Leinweber:1998ej,Hackett-Jones:2000qk,Hemmert:2002uh,Young:2004tb}.
However, there is no model independent way to respect the constraints of
chiral symmetry over the range of $Q^{2}$ and $m_{\pi}$ required by the QCDSF
data. Instead, at finite $Q^{2}$, one has been led to study various
phenomenological parameterizations~\cite{Ashley:2003sn}, which have at least
ensured the correct leading order non-analytic structure as $m_{\pi
}\rightarrow0$. Our purpose here is three-fold. First, we wish to use the
lattice data to investigate whether a particular quark model is capable of
describing the properties of the nucleon in this additional dimension of
varying $m_{\pi}$ - an important test which any respectable quark model should
satisfy\footnote{Just as the study of QCD as a function of $N_{c}$ has proven
extremely valuable, so the study of hadron properties as a function of quark
mass, using the results of lattice QCD calculations, undoubtedly offers
significant \ insight into QCD, as well as new ways to model it
\cite{Cloet:2002eg}.}. Second, having confirmed that the model is consistent
with the lattice data over the range of $m_{\pi}$ noted earlier, we use the
model to extrapolate to large values of $Q^{2}$ (for lattice values of
$m_{\pi}$). Third, we also use the model to extrapolate to the physical pion mass.

The model which we consider here is the light front cloudy bag model
(LFCBM)~\cite{Miller:2002ig}, which was developed as a means of preserving the
successes of the original cloudy bag model \cite{cbm}, while ensuring
covariance in order to deal unambiguously with modern high energy experiments.
The light front constituent quark model, upon which it is built
\cite{Frank:1995pv}, predicted the rapid decrease of $G_{E}/G_{M}$ with
$Q^{2}$ and, as the pion cloud is expected to be relatively unimportant at
large $Q^{2}$, this success carries over to the LFCBM \cite{Miller:2002ig}.
Furthermore the LFCBM corresponds to a Lagrangian built upon chiral symmetry,
so it can be extended to the limit of low quark mass as well as low and high
$Q^{2}$.

The outline of the paper follows. In Sect. 2 we briefly review the LFCBM. In
Sect. 3 we present the lattice QCD data, explain the fitting procedure and
present the results. Sect. 4 contains some concluding remarks.

\section{Review of the LFCBM}

The light front cloudy bag model (LFCBM) respects chiral symmetry and Lorentz
invariance and reproduces the four nucleon electromagnetic form factors.
Therefore it is reasonable to try to use it to extrapolate the form factors
computed using lattice QCD to the physical pion mass. We begin by briefly
introducing the key features of the LFCBM.

The LFCBM is a \textit{relativistic} constituent quark model incorporating the
effect of \textit{pion-loops}, key features motivated by chiral symmetry. The
light-front dynamics is employed to maintain the Poincar\'{e} invariance, and
one pion-loop corrections are added to incorporate significant pion cloud
effects (particularly in the neutron electric form factor and magnetic
moments) as well as the leading non-analytic behavior imposed by chiral
symmetry. In light-front dynamics the fields are quantized at a fixed
\textquotedblleft time\textquotedblright$=\tau=x^{0}+x^{3}\equiv x^{+}.$ The
light front time or $\tau$-development operator is then $P^{0}-P^{3}\equiv
P^{-}$. The canonical spatial variable is $x^{-}=x^{0}-x^{3}$, with a
canonical momentum $P^{+}=P^{0}+P^{3}$. The other coordinates are
$\mathbf{x}_{\perp}$ and $\mathbf{P}_{\perp}$. The relation between the energy
and momentum of a free particle is given by $p^{-}=(p_{\perp}^{2}+m^{2}%
)/p^{+}$, with the quadratic form allowing the separation of center of mass
and relative coordinates. The resulting wave functions are frame independent.
The light front technique is particularly relevant for calculating form
factors because one uses boosts that are independent of interactions.

Our goal is to calculate the Dirac $F_{1}$ and Pauli $F_{2}$ form factors
given by:%
\begin{equation}
\left\langle N,\lambda^{\prime}p^{\prime}\left\vert J^{\mu}\right\vert
N,\lambda p\right\rangle =\overline{u}_{\lambda^{\prime}}(p^{\prime})\left[
F_{1}(Q^{2})\gamma^{\mu}+\frac{F_{2}(Q^{2})}{2M_{N}}i\sigma^{\mu\nu}%
(p^{\prime}-p)_{\nu}\right]  u_{\lambda}(p)\,. \label{FORMULA_EM_CURRENT}%
\end{equation}
The momentum transfer is $q^{\mu}=($ $p^{\prime}-p)^{\mu},~Q^{2}=-q^{2}$ and
$J^{\mu}$ is taken to be the electromagnetic current operator for a free
quark. For $Q^{2}=0$ the form factors $F_{1}$ and $F_{2}$ are, respectively,
equal to the charge and the anomalous magnetic moment $\kappa$ in units of $e$
and $e/(2M_{N})$ , and the magnetic moment is $\mu=F_{1}(0)+F_{2}(0)=1+\kappa
$. The evaluation of the form factors is simplified by using the so-called
Drell-Yan reference frame in which $q^{+}=0$, so that $Q^{2}=q_{\bot}%
^{2}=q_{1}^{2}$. If light-front spinors for the nucleons are used, the form
factors can be expressed in terms of matrix elements of the plus component of
the current \cite{Brodsky:1980zm}:%
\begin{equation}
F_{1}(Q^{2})=\langle N,\uparrow\left\vert J^{+}\right\vert N,\uparrow
\rangle,\;\hspace{1cm}\mathrm{and}\hspace{1cm}QF_{2}(Q^{2})=(-2M_{N})\langle
N,\uparrow\left\vert J^{+}\right\vert N,\downarrow\rangle\,.
\label{FORMULA_F1_F2_CURRENT}%
\end{equation}
The form factors are calculated using the \textquotedblleft
good\textquotedblright\ component of the current, $J^{+}$, to suppress the
effects of quark-pair terms. Finally, we note that in our fits we will use the
Sachs form factors, which are defined as
\begin{equation}
G_{E}=F_{1}-\frac{Q^{2}}{4M_{N}^{2}}F_{2},\hspace{2cm}G_{M}=F_{1}+F_{2}
\label{FORMULA_G_TO_F}%
\end{equation}

The next step is to construct the bare (pionless) nucleon wave function $\Psi
$, which is a symmetric function of the quark momenta, independent of
reference frame, and an eigenstate of the canonical spin operator. The
commonly used ansatz is:
\begin{align}
\Psi(p_{i})  &  =\Phi(M_{0}^{2})u(p_{1})u(p_{2})u(p_{3})\psi(p_{1},p_{2}%
,p_{3}),\label{FORMULA_WAV_FUNC}\\
p_{i}  &  =\mathbf{p}_{i}s_{i},~\tau_{i},\nonumber
\end{align}
where $\psi$ is a spin-isospin color amplitude factor, the $p_{i}$ are
expressed in terms of relative coordinates, the $u(p_{i})$ are Dirac spinors
and $\Phi$ is a momentum distribution wave function. The specific form of
$\psi$ is given in Eq.~(12) of Ref.~\cite{Miller:2002qb} and earlier in Ref.
\cite{Chung:1991st}. This is a relativistic version of the familiar SU(6) wave
function, with no configuration mixing included. The notation is that
$\mathbf{p}_{i}=(p_{i}^{+},$ $\mathbf{p}_{i\bot})$. The total momentum
is\ $\mathbf{P}=\mathbf{p}_{1}+\mathbf{p}_{2}+\mathbf{p}_{3}$, the relative
coordinates are $\xi=p_{1}^{+}/(p_{1}^{+}+p_{2}^{+})$,~$\eta=(p_{1}^{+}%
+p_{2}^{+})/P^{+}$, and $\mathbf{k}_{\bot}=(1-\xi)\mathbf{p}_{1\bot}%
-\xi\mathbf{p}_{2\bot}$,~$\mathbf{K}_{\bot}=(1-\eta)(\mathbf{p}_{1\bot
}+\mathbf{p}_{2\bot})-\eta\mathbf{p}_{3\bot}$. In computing a form factor, we
take quark 3 to be the one struck by the photon. The value of $1-\eta$ is not
changed $(q^{+}=0)$, so only one relative momentum, $\mathbf{K}_{\bot}$ is
changed: $\mathbf{K}_{\bot}^{\prime}=\mathbf{K}_{\bot}-\eta\mathbf{q}_{\bot}$.
The form of the momentum distribution wave function is taken from Schlumpf
\cite{Schlumpf:1992ce}:%
\begin{equation}
\Phi(M_{0})={N}{\left(  M_{0}^{2}+\beta^{2}\right)  ^{\gamma}}\,,
\label{FORMULA_SPAT_WAV_FUNC}%
\end{equation}
with $M_{0}^{2}$ the mass-squared operator for a non-interacting system:%
\begin{equation}
M_{0}^{2}=\frac{K_{\bot}^{2}}{\eta(1-\eta)}+\frac{k_{\bot}^{2}+M^{2}}{\eta
\xi(1-\xi)}+\frac{M^{2}}{1-\eta}\,. \label{FORMULA_M0}%
\end{equation}
Schlumpf's parameters were $\beta=0.607~%
\operatorname{GeV}%
$, $\gamma=-3.5$, $M=0.267~%
\operatorname{GeV}%
$, where the value of $\gamma$ was chosen so that $Q^{4}G_{M}(Q^{2})$ is
approximately constant for $Q^{2}>4~%
\operatorname{GeV}%
^{2}$, in accord with experimental data. The parameter $\beta$ helps govern
the values of the transverse momenta allowed by the wave function $\Phi$ and
is closely related to the rms charge radius. The constituent quark mass, $M$,
was primarily determined by the magnetic moment of the proton. We shall use
different values when including the pion cloud and fitting lattice data.

\begin{figure}
[ptb]
\begin{center}
\includegraphics[
height=1.6734in,
width=5.2529in
]%
{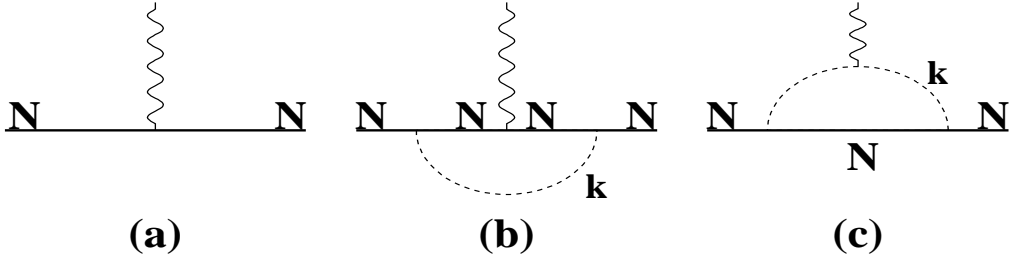}%
\caption{Diagrams}%
\label{fig:diagrams}%
\end{center}
\end{figure}

A physical nucleon can sometimes undergo a quantum fluctuation so that it
consists of a bare nucleon and a virtual pion. In this case, an incident
photon can interact electromagnetically with a bare nucleon,
Fig.~\ref{fig:diagrams}a, with a nucleon while a pion is present,
Fig.~\ref{fig:diagrams}b, or with a charged pion in flight,
Fig.~\ref{fig:diagrams}c. These effects are especially pronounced for the
neutron $G_{E}$\cite{cbm}, at small values of $Q^{2}$. The tail of the
negatively charged pion distribution extends far out into space, causing the
mean square charge radius, $R_{n}^{2}$, to be negative. The effects of the
pion cloud need to be computed relativistically if one is to confront data
taken at large $Q^{2}$. This involves evaluating the Feynman diagrams of
Fig.~\ref{fig:diagrams} using photon-bare-nucleon form factors from the
relativistic model, and using a relativistic $\pi$-nucleon form factor. The
resulting model is defined as the light-front cloudy bag model LFCBM
\cite{Miller:2002ig}. The light-front treatment is implemented by evaluating
the integral over the virtual pion four-momentum $k^{\pm},\mathbf{k}_{\perp}$,
by first performing the integral over $k^{-}$ analytically, re-expressing the
remaining integrals in terms of relative variables ($\alpha=k^{+}/p^{+})$, and
shifting the relative $_{\perp}$ variable to $\mathbf{L}_{\perp}$ to simplify
the numerators. Thus the Feynman graphs, Fig.~\ref{fig:diagrams}, are
represented by a single $\tau$-ordered diagram. The use of $J^{+}$ and the Yan
identity\cite{Chang:qi} $S_{F}(p)=\sum_{s}u(p,s)\overline{u}(p,s)/(p^{2}%
-m^{2}+i\epsilon)+\gamma^{+}/2p^{+}$ allows one to see that the nucleon
current operators appearing in Fig.\ref{fig:diagrams}b act between
on-mass-shell spinors.%

The results can be stated as
\begin{equation}
F_{i\alpha}(Q^{2})=Z\left[  F_{i\alpha}^{(0)}(Q^{2})+F_{ib\alpha}%
(Q^{2})+F_{ic\alpha}(Q^{2})\right]  , \label{FORMULA_F_TO_F0_FB_FC}%
\end{equation}
where $i=1,2$ denotes the Dirac and Pauli form factors, $\alpha=n,p$
determines the identity of the nucleon, and $F_{i\alpha}^{(0)}(Q^{2})$ are the
form factors computed in the absence of pionic effects. The wave function
renormalization constant, $Z$, is determined from the condition that the
charge of the proton be unity: $F_{1p}(Q^{2}=0)=1$. For illustration we start
with the calculation of the neutron form factors. Then, evaluating the graph
in {}Fig.~\ref{fig:diagrams}b gives%
\begin{multline}
F_{1bn}(Q^{2})=g_{0}^{2}\int_{0}^{1}d\alpha\alpha\int{\frac{d^{2}L}{(2\pi
)^{3}}}R_{N}({\mathbf{L}^{(+)}}^{\;2},\alpha)R_{N}({\mathbf{L}^{(-)}}%
^{\;2},\alpha)\label{FORMULA_F1BN}\\
\lbrack(F_{1p}^{({0})}(Q^{2})+F_{1n}^{({0})}(Q^{2})/2)(\alpha^{2}(M^{2}%
-Q^{2}/4)+L^{2})-(F_{2p}^{({0})}(Q^{2})+F_{2n}^{({0})}(Q^{2})/2))(\alpha
^{2}Q^{2}/2)],
\end{multline}%
\begin{multline}
F_{2bn}(Q^{2})=-g_{0}^{2}\int_{0}^{1}d\alpha\alpha\int{\frac{d^{2}L}%
{(2\pi)^{3}}}R_{N}({\mathbf{L}^{(+)}}^{\;2},\alpha)R_{N}({\mathbf{L}^{(-)}%
}^{\;2},\alpha)\label{FORMULA_F2BN}\\
\lbrack(F_{1p}^{({0})}(Q^{2})+{\frac{1}{2}}F_{1n}^{({0})}(Q^{2}))(2\alpha
^{2}M^{2})\\
+(F_{2p}^{({0})}(Q^{2})+{\frac{1}{2}}F_{2n}^{({0})}(Q^{2})))(\alpha^{2}%
M^{2}(1-Q^{2}/4M^{2})+(L_{x}^{2}-L_{y}^{2}))]
\end{multline}
where $g_{0}$ is the bare $\pi$N coupling constant, and the renormalized
coupling constant $Zg_{0}^{2}=g^{2}/4\pi=13.5\;$,$\mathbf{L}_{\perp}^{(\pm
)}\equiv\mathbf{L}_{\perp}\pm\alpha\mathbf{q}_{\perp}/2$, $\alpha\equiv
{\frac{k^{+}}{p^{+}}}$, $D_{N}(k_{\perp}^{2},\alpha)\equiv{M^{2}\alpha
^{2}+k_{\perp}^{2}+\mu^{2}(1-\alpha)}$, and $R_{N}(k_{\perp}^{2},\alpha
)\equiv{\frac{F_{\pi N}^{N}(k_{\perp}^{2},\alpha)}{D_{N}(k_{\perp}^{2}%
,\alpha)}}.$ The $\pi$N form factor is taken as
\cite{Zoller:1991cb,Szczurek:gw}
\begin{equation}
F_{\pi N}(k_{\perp}^{2},\alpha)=e^{{-(D_{N}(k_{\perp}^{2},\alpha
)/2\alpha(1-\alpha)\Lambda^{2})}}{,} \label{FORMULA_PI_N_FF}%
\end{equation}
and maintains charge conservation \cite{Speth:pz}. The constant $\Lambda$ is a
free parameter, but very large values are excluded by the small flavor
asymmetry of the nucleon sea.

{}From Eqns.~(\ref{FORMULA_F1BN}) and (\ref{FORMULA_F2BN}) we see that each
term in the nucleon current operator contributes to both $F_{1}$ and $F_{2}$.
The evaluation of graph \ref{fig:diagrams}c yields
\begin{equation}
F_{1cn}(Q^{2})=-g_{0}^{2}F_{\pi}(Q^{2})\int_{0}^{1}d\alpha\alpha\int
{\frac{d^{2}K}{(2\pi)^{3}}}R({\mathbf{K}^{(+)}}^{\;2},\alpha)R({\mathbf{K}%
^{(-)}}^{\;2},\alpha)\left[  K^{2}+M^{2}\alpha^{2}-(1-\alpha)^{2}{\frac{Q^{2}%
}{4}}\right]  \label{FORMULA_F1CN}%
\end{equation}%
\begin{equation}
F_{2cn}(Q^{2})=-g_{0}^{2}(2M^{2})F_{\pi}(Q^{2})\int_{0}^{1}d\alpha\alpha
^{2}(1-\alpha)\int{\frac{d^{2}K}{(2\pi)^{3}}}R({\mathbf{K}^{(+)}}^{\;2}%
,\alpha)R({\mathbf{K}^{(-)}}^{\;2},\alpha) \label{FORMULA_F2CN}%
\end{equation}
where $\mathbf{K}_{\perp}^{(\pm)}\equiv\mathbf{K}_{\perp}\pm(1-\alpha
)\mathbf{q}_{\perp}/2$.\footnote{These formulae are slightly different from
those of Ref.~\cite{Miller:2002ig}. This leads to slight changes in the
parameters that will be discussed elsewhere.}

The proton form factors can be obtained by simply making the replacements
$n\rightarrow p$ in Eqs.~(\ref{FORMULA_F1BN},\ref{FORMULA_F2BN}) and
$-g_{0}^{2}\rightarrow g_{0}^{2}$ in Eqs.~(\ref{FORMULA_F1CN}%
,\ref{FORMULA_F2CN}). The change in sign accounts for the feature that the
$\pi^{-}$ cloud of the neutron becomes a $\pi^{+}$ cloud for the proton. The
mean-square isovector radii $\left\langle r^{2}\right\rangle _{i}^{V}$,
computed using Eqs.~(\ref{FORMULA_F_TO_F0_FB_FC}), and then taken to the
chiral limit at low-$Q^{2}$, have the same singular $\log$ terms as those of
the relativistic results of Beg and Zepeda \cite{Beg:1973sc}.

The LFCBM was defined by choosing four free parameters: $m,\beta
,\gamma,\Lambda$ so as to best reproduce the four experimentally measured
electromagnetic form factors of the nucleon \cite{Miller:2002ig}. In the
present work, the most relevant of these parameters will be varied to
reproduce lattice data, and the resulting dependence on the quark mass and
lattice spacing used to extrapolate to the physical region.

\section{Fitting the QCDSF Form Factors and Extrapolating to the Physical Pion
Mass.}

In this section we discuss the fitting procedure used to parametrize the
nucleon form factors calculated in lattice QCD. We use data produced by the
QCDSF Collaboration \cite{QCDSF Collaboration} and employ the LFCBM to
calculate the corresponding form factors, varying the model parameters to find
the best-fit to the different sets of lattice data obtained for different
values of the current quark mass, $m_{q}$. The behavior of the fitting
parameters is then represented by a polynomial function of the quark mass
$m_{q}$. This polynomial fit in $m_{q}$, or equivalently in pion mass squared,
$m_{\pi}^{2}$, can then be used to extrapolate the values of the fitting
parameters to the physical pion mass. Nucleon form factors for the physical
pion mass are then calculated using the extrapolated values for the model
parameters. In the following few subsections a more elaborate explanation is
given and the results are presented. In section \ref{QCDSF_DATA}\ we describe
the available data and the analysis procedure used to extract the quantities
necessary for further fits. In sec. \ref{DATA_FIT} we describe the details of
the fitting and extrapolation process and in the sec. \ref{M_PHYS} we present
the nucleon form factors resulting from the extrapolation to the physical pion
mass and make comparisons with experiment.

\subsection{QCDSF Data and Its Analysis\label{QCDSF_DATA}}

The form factor calculations in Ref.\cite{QCDSF Collaboration} \ were carried
out for three different values of the lattice spacing, $a=\{0.47,0.34,0.26\}%
\operatorname{GeV}%
^{-1}$. For each value of $a$ several sets of pion (or equivalently nucleon)
masses were considered. For each mass set Dirac and Pauli form factors for
both the proton and neutron were calculated at several values of $Q^{2}$. The
typical range for the pion mass used varied from $1.2%
\operatorname{GeV}%
$ to $0.6%
\operatorname{GeV}%
$, with the corresponding nucleon mass ranging from approximately $2%
\operatorname{GeV}%
$ to $1.5%
\operatorname{GeV}%
$. The typical range for $Q^{2}$ was $0.6%
\operatorname{GeV}%
^{2}$ to $2.3%
\operatorname{GeV}%
^{2}$.

The LFCBM is basically a relativistic constituent quark model, so we need to
relate the model constituent mass of Eq.~(\ref{FORMULA_M0}) to the masses of
the nucleon and pion. To do so we use the approach of Ref.~\cite{Cloet:2002eg}%
, ( Eq.~(8))%
\begin{equation}
M=M_{\chi}+\frac{cm_{q}^{phys}}{(m_{\pi}^{phys})^{2}}m_{\pi}^{2},
\label{MCHI_EQ}%
\end{equation}
where $M_{\chi}$\ is the constituent quark mass in the chiral limit,
$m_{q}^{phys}$\ is the current quark mass and $c$ is of order 1. In the study
of octet magnetic moments in the AccessQM model of Ref.~\cite{Cloet:2002eg},
the best fit value for $M_{\chi}$ was $0.42%
\operatorname{GeV}%
$, while for $cm_{q}^{phys}$ it was $0.0059%
\operatorname{GeV}%
$.

\subsection{Lattice Data Fit and Extrapolation\label{DATA_FIT}}

The first step in our extrapolation of the lattice results to the physical
quark mass is to fit the lattice results for each quark mass $m_{q}$ by
adjusting the parameters of the LFCBM calculation. For that purpose two
fitting parameters were chosen. The first parameter is $M_{\chi}$ in
Eq.~(\ref{MCHI_EQ}), which determines the constituent quark mass. This
parameter was varied for each lattice spacing separately, since some
dependence upon lattice spacing was anticipated. The second parameter is the
internal parameter, $\gamma$, in the nucleon wave function Eq.
(\ref{FORMULA_SPAT_WAV_FUNC}), which is varied separately for each pion (or
equivalently nucleon) mass. For convenience, we express all magnetic form
factors \textbf{$G_{M}$ }in \textquotedblleft physical" units of
$e/2M_{N}^{Physical}$. Since the LFCBM\ uses the mass of the $\rho$-meson
included in the pion electromagnetic form factor, we need the extrapolated
value for its mass. We use the simple fitting function from
Ref.~\cite{RHO_MASS}:%
\begin{equation}
m_{\rho}=c_{0}+c_{1}m_{\pi}^{2}, \label{M_RHO}%
\end{equation}
with $c_{0}=0.776%
\operatorname{GeV}%
$ and $c_{1}=0.427%
\operatorname{GeV}%
^{-1}$ .

\begin{figure}
[ptb]
\begin{center}
\includegraphics[
height=4.5636in,
width=4.9009in
]%
{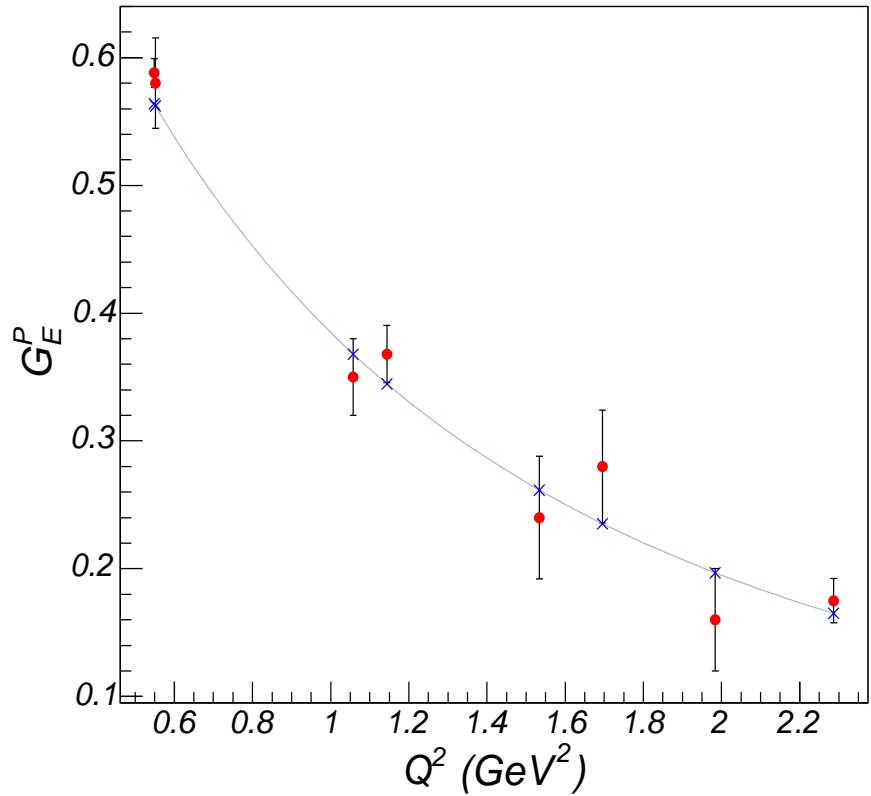}%
\caption{(Color online) LFCBM fit to QCDSF data for $G_{E}^{P}(in~units~of~e)$
for a lattice spacing $a=0.26~\operatorname{GeV}^{-1}$, $M_{P}%
=1.80~\operatorname{GeV}$ and $m_{\pi}=0.93~\operatorname{GeV}$}%
\label{PLOT_PROTON_GE_6_4_3}%
\end{center}
\end{figure}

\begin{figure}
[ptbptb]
\begin{center}
\includegraphics[
height=4.5455in,
width=4.8836in
]%
{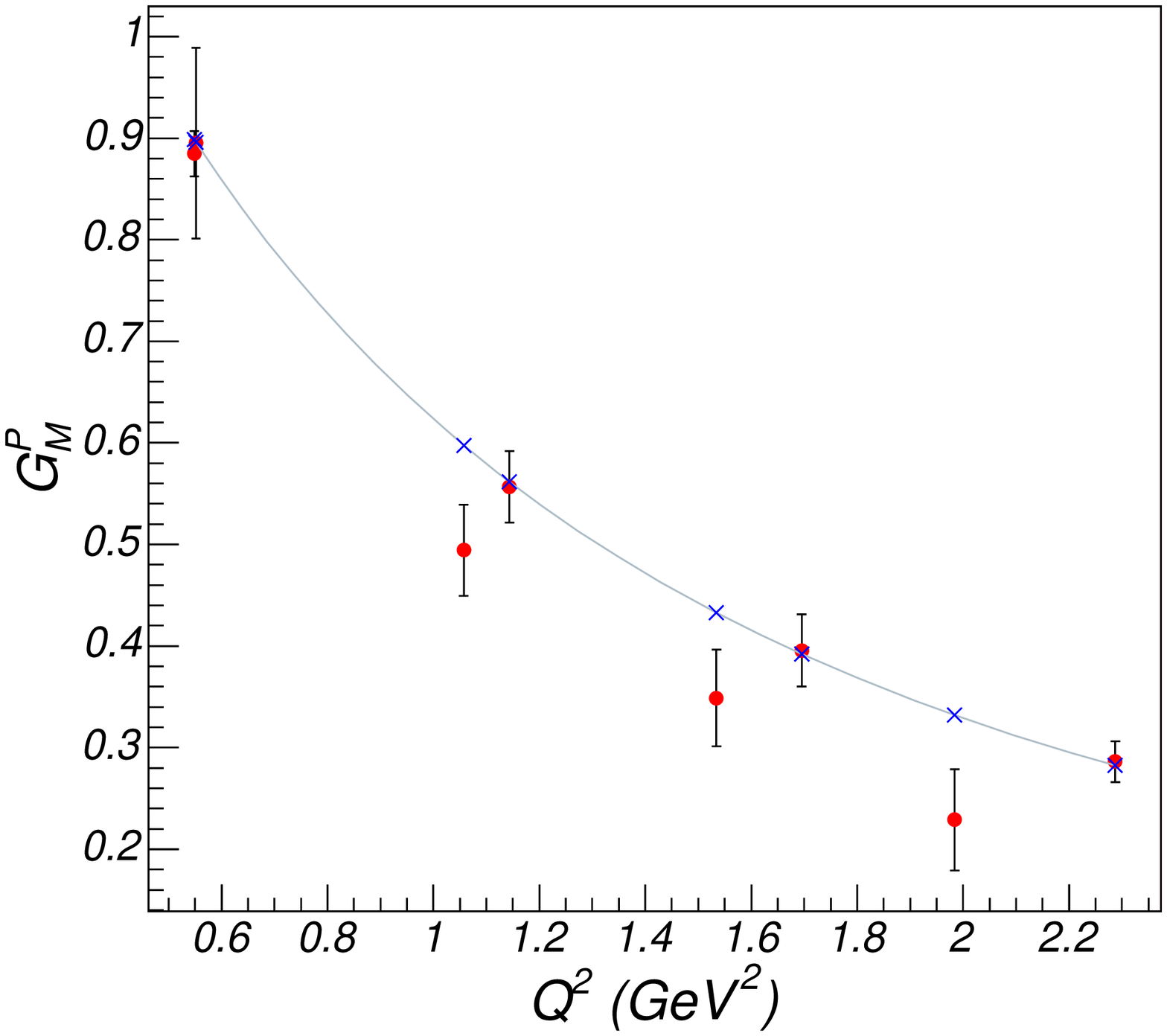}%
\caption{(Color online) LFCBM fit to QCDSF data for $G_{M}^{P}%
(in~units~of~e/(2M_{N}^{Physical}))$ for a lattice spacing
$a=0.26~\operatorname{GeV}^{-1}$, $M_{P}=1.80~\operatorname{GeV}$ and $m_{\pi
}=0.93~\operatorname{GeV}$}%
\label{PLOT_PROTON_GM_6_4_3}%
\end{center}
\end{figure}

\begin{figure}
[ptbptbptb]
\begin{center}
\includegraphics[
height=4.5636in,
width=4.9009in
]%
{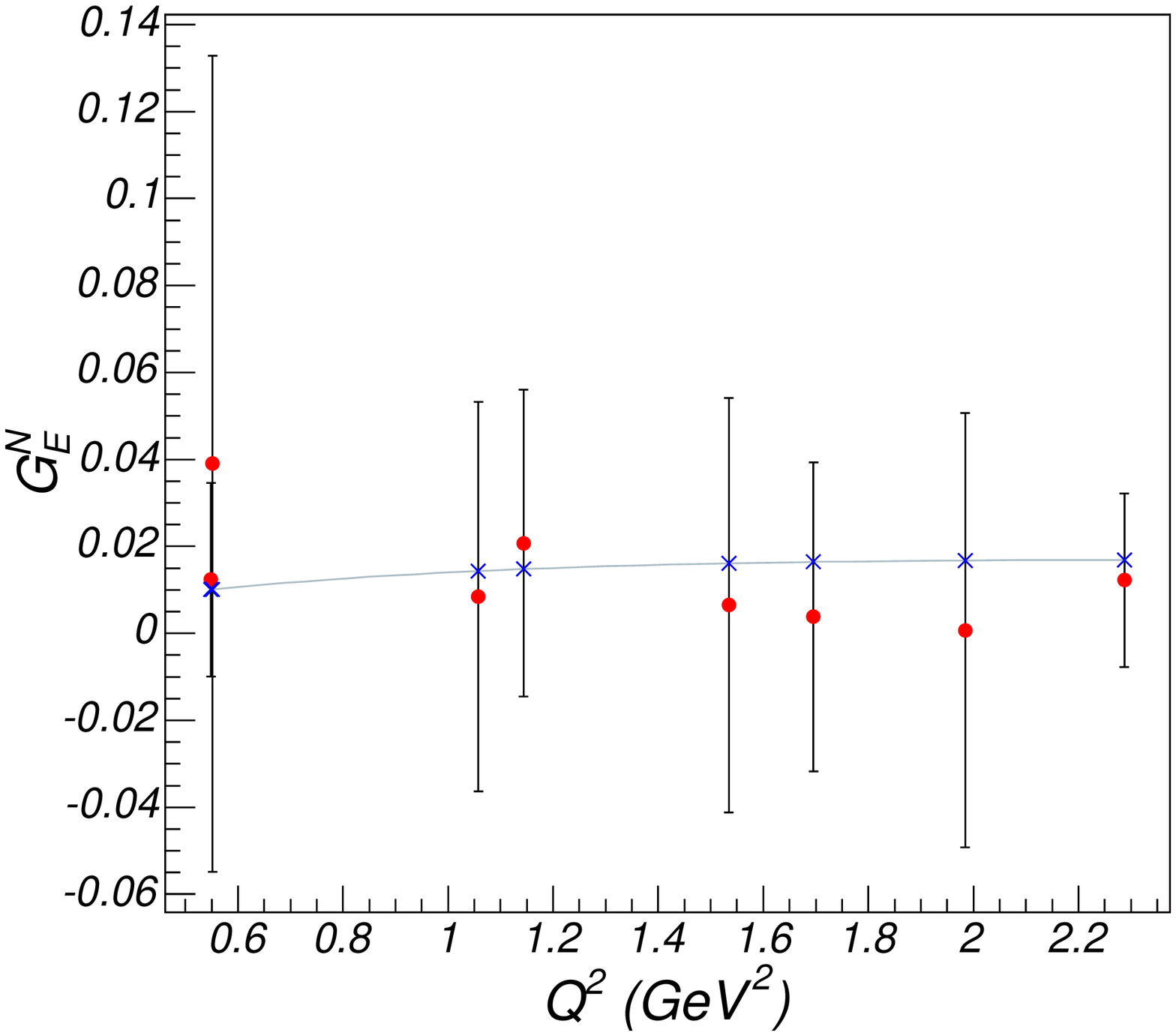}%
\caption{(Color online) LFCBM fit to QCDSF data for $G_{E}^{N}(in~units~of~e)$
for a lattice spacing $a=0.26~\operatorname{GeV}^{-1}$, $M_{P}%
=1.80~\operatorname{GeV}$ and $m_{\pi}=0.93~\operatorname{GeV}$}%
\label{PLOT_NEUTRON_GE_6_4_3}%
\end{center}
\end{figure}

\begin{figure}
[ptbptbptbptb]
\begin{center}
\includegraphics[
height=4.5455in,
width=4.8836in
]%
{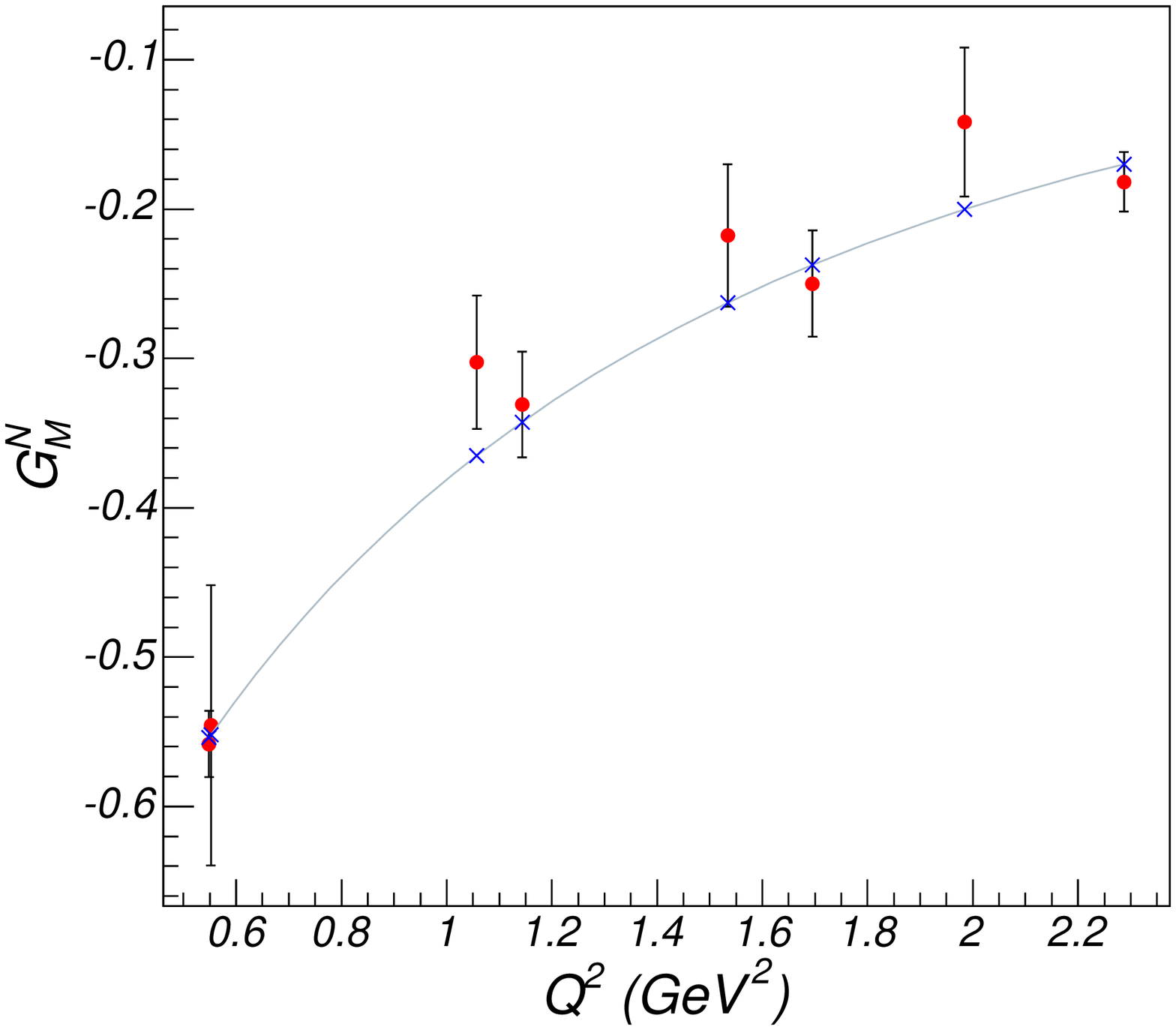}%
\caption{(Color online) LFCBM fit to QCDSF data for $G_{M}^{N}%
(in~units~of~e/(2M_{N}^{Physical}))$ for a lattice spacing
$a=0.26~\operatorname{GeV}^{-1}$, $M_{P}=1.80~\operatorname{GeV}$ and $m_{\pi
}=0.93~\operatorname{GeV}$}%
\label{PLOT_NEUTRON_GM_6_4_3}%
\end{center}
\end{figure}

A function representing the $\chi^{2}$ for the deviation between the lattice
data and the values calculated using the LFCBM was constructed and minimized
by varying the fitting parameters. Changing the value of $M_{\chi}$ causes the
the calculated form factors to move up or down by an amount approximately
independent of $Q^{2}$, thereby causing a relatively small change in $\chi
^{2}$. Therefore a simple grid variation for that parameter was employed, with
grid boundaries $M_{\chi}\in\left[  0.15,0.45\right]
\operatorname{GeV}%
$, and step size of $\delta_{M_{\chi}}=0.01%
\operatorname{GeV}%
$. As for the parameter, $\gamma$, the variation of $\chi^{2}$ was much
stronger and the Minuit package of CERN's Root framework \cite{ROOT} was used
for the minimization. At first the boundaries for $\gamma$ were set to keep it
in the physical region, but successful boundless runs were also performed in
order to confirm the true minimum and error sizes. The pion masses used in the
lattice calculation are very large, and the resulting pionic effects are very
small. Therefore the value of $\Lambda$ could not be determined from lattice
data and its value was held fixed at $\Lambda=0.58%
\operatorname{GeV}%
$ Similarly, varying $\beta$ did not change the description of the lattice
data, so it was held fixed at $\beta=0.607%
\operatorname{GeV}%
/c$. The resulting fits are in good agreement with data, as one can see in
Figs. \ref{PLOT_PROTON_GE_6_4_3}, \ref{PLOT_PROTON_GM_6_4_3},
\ref{PLOT_NEUTRON_GE_6_4_3} and \ref{PLOT_NEUTRON_GM_6_4_3}. The best-fit
values of the parameters are shown in Table I. The figures show results for
the smallest lattice spacing, $a=0.26%
\operatorname{GeV}%
^{-1}$, but the reproduction of lattice data is equally successful for larger
values of $a$.
%

\begin{gather*}
\text{TABLE I.~Lattice data and LFCBM fitting parameters.(All expressed in
powers of }%
\operatorname{GeV}%
\text{.)}\\%
\begin{tabular}
[c]{p{0.8in}p{0.8in}p{0.8in}p{0.8in}p{0.8in}}\hline\hline
$a$ & $m_{\pi}$ & $M_{N}$ & $M_{\chi}$ & $\gamma$\\\hline\hline
$0.47$ & $1.146$ & $2.062$ & $0.390(5)$ & $-6.12(7)$\\
$0.47~$ & $1.068$ & $1.981$ & $0.390(5)$ & $-5.67(6)$\\
$0.47$ & $0.873$ & $1.746$ & $0.390(5)$ & $-4.95(9)$\\
$0.47$ & $0.752$ & $1.567$ & $0.390(5)$ & $-4.78(12)$\\
$0.47$ & $0.638$ & $1.503$ & $0.390(5)$ & $-4.67(15)$\\
$\mathit{0.47}$ & $\mathit{0.135}$ & $\mathit{0.938}$ & $\mathit{0.390(5)}$ &
$\mathit{-4.79(46)}$\\\hline
$0.34$ & $1.201$ & $2.141$ & $0.280(5)$ & $-5.03(6)$\\
$0.34$ & $1.035$ & $1.933$ & $0.280(5)$ & $-4.37(5)$\\
$0.34$ & $0.881$ & $1.732$ & $0.280(5)$ & $-4.99(5)$\\
$0.34$ & $0.706$ & $1.522$ & $0.280(5)$ & $-3.51(6)$\\
$\mathit{0.34}$ & $\mathit{0.135}$ & $\mathit{0.938}$ & $\mathit{0.280(5)}$ &
$\mathit{-2.91(29)}$\\\hline
$0.26$ & $1.237$ & $2.202$ & $0.210(5)$ & $-4.78(5)$\\
$0.26$ & $1.092$ & $2.028$ & $0.210(5)$ & $-4.14(7)$\\
$0.26$ & $0.925$ & $1.802$ & $0.210(5)$ & $-3.69(5)$\\
$0.26$ & $0.744$ & $1.600$ & $0.210(5)$ & $-3.09(6)$\\
$0.26$ & $0.580$ & $1.379$ & $0.210(5)$ & $-3.01(13)$\\
$\mathit{0.26}$ & $\mathit{0.135}$ & $\mathit{0.938}$ & $\mathit{0.210(5)}$ &
$\mathit{-2.41(22)}$\\\hline
\end{tabular}
\end{gather*}

\begin{figure}
[ptb]
\begin{center}
\includegraphics[
height=3.5016in,
width=3.6772in
]%
{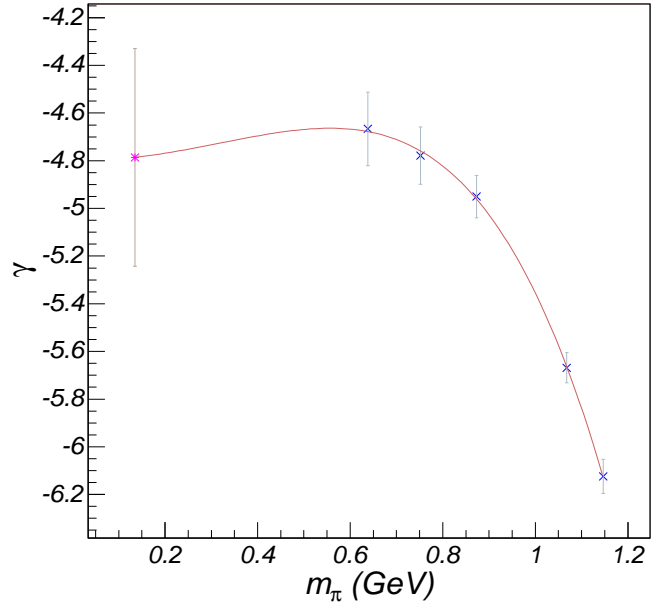}%
\caption{(Color online) Polynomial extrapolation of $\gamma$ vs. $m_{\pi}$ for
lattice spacing $a=0.47~GeV^{-1}$ }%
\label{PLOT_GAMMA_VS_MPI_6_0}%
\end{center}
\end{figure}

\begin{figure}
[ptbptbh]
\begin{center}
\includegraphics[
height=3.5129in,
width=3.6884in
]%
{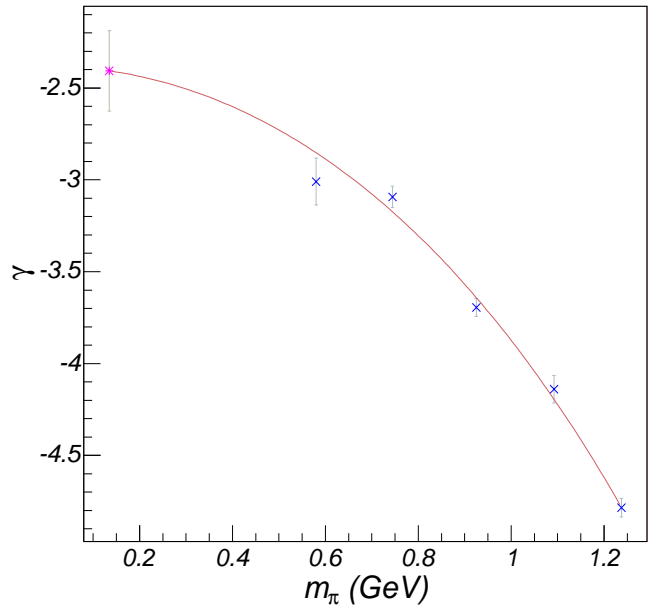}%
\caption{(Color online) Polynomial extrapolation of $\gamma$ vs. $m_{\pi}$ for
lattice spacing $a=0.26~GeV^{-1}$}%
\label{PLOT_GAMMA_VS_MPI_6_4}%
\end{center}
\end{figure}

The next step is to extrapolate the fitting parameters to the physical quark
mass. This is done using the assumption that the parameters vary smoothly as
functions of the quark mass, and the fact that $m_{q}\thicksim m_{\pi}^{2}$
over the mass range investigated. We limited the extrapolation function to a
low order polynomial in $m_{\pi}^{2}$. The resulting fits for two lattice
spacings are presented in Figs. \ref{PLOT_GAMMA_VS_MPI_6_0} and
\ref{PLOT_GAMMA_VS_MPI_6_4}, from which we see that the fitting function
provides a very accurate representation of the values obtained from lattice
data.%
The fitted values of $\gamma$ and the extrapolation to the physical value of
$m_{\pi}$, with their corresponding errors, are shown in Figs.
\ref{PLOT_GAMMA_VS_MPI_6_0} and \ref{PLOT_GAMMA_VS_MPI_6_4}.

\begin{figure}
[ptb]
\begin{center}
\includegraphics[
height=4.4564in,
width=4.6899in
]%
{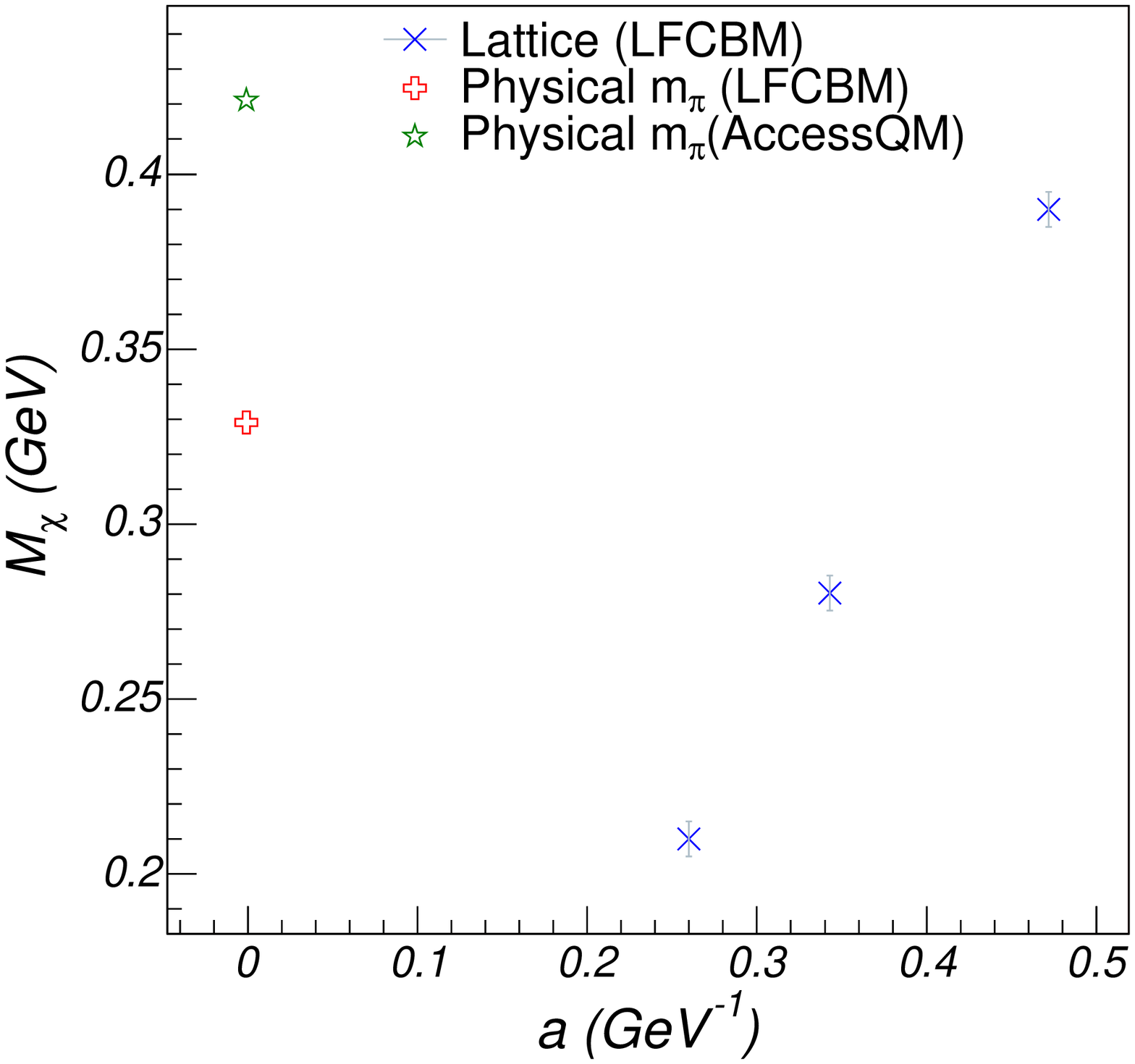}%
\caption{(Color online) Variation of $M_{\chi}$with lattice spacing $a.$The
best-fit values of $M_{\chi}$ for physical $m_{\pi}$ using LFCBM and AcessQM
models are presented as well.}%
\label{PLOT_MCHI_VS_LAT_A}%
\end{center}
\end{figure}

In our calculations, $M_{\chi}$ has a very weak dependence on the pion mass,
but it has a rather strong dependence upon the lattice spacing. As we see in
Table~I and Figs. \ref{PLOT_PROTON_GE_6_4_3}-\ref{PLOT_NEUTRON_GM_6_4_3}, very
good fits to the lattice data are obtained even without varying $M_{\chi}$ for
each quark mass. By contrast, Fig. \ref{PLOT_MCHI_VS_LAT_A} and Table I show
rather dramatic variation of $M_{\chi}$ for different values of the lattice
spacing, $a$. This suggests that the larger values of the lattice spacing are
rather far from the continuum limit and (at best) only the results for the
smallest lattice spacing should be compared with experimental data. It would
clearly be desirable to have new data at even smaller $a$, or using an
improved action, known to provide a good approximation to the continuum limit.%

\begin{figure}
[ptb]
\begin{center}
\includegraphics[
height=7.8395in,
width=6.3235in
]%
{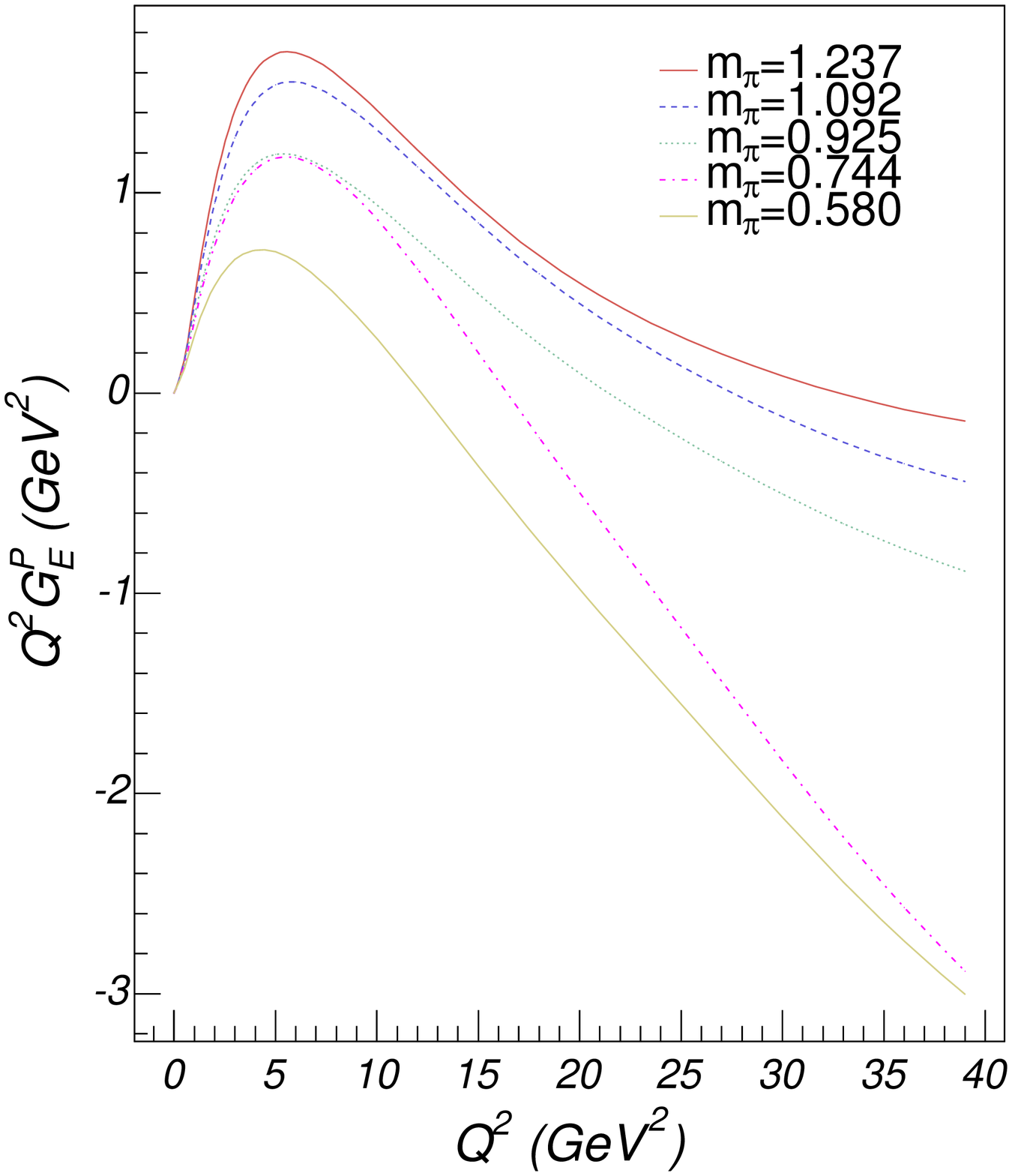}%
\caption{(Color online) LFCBM calculations using parameters (Figs.
\ref{PLOT_GAMMA_VS_MPI_6_0}, \ref{PLOT_GAMMA_VS_MPI_6_4} and
\ref{PLOT_MCHI_VS_LAT_A}) obtained by fitting the lattice results for the
proton electric form factor, $G_{E}$, at lattice \ spacing $a=0.26~GeV^{-1}$.}%
\label{PLOT_COMPARE_LAT_PROT_GE_6_4}%
\end{center}
\end{figure}

\begin{figure}
[ptbptb]
\begin{center}
\includegraphics[
height=7.8395in,
width=6.3235in
]%
{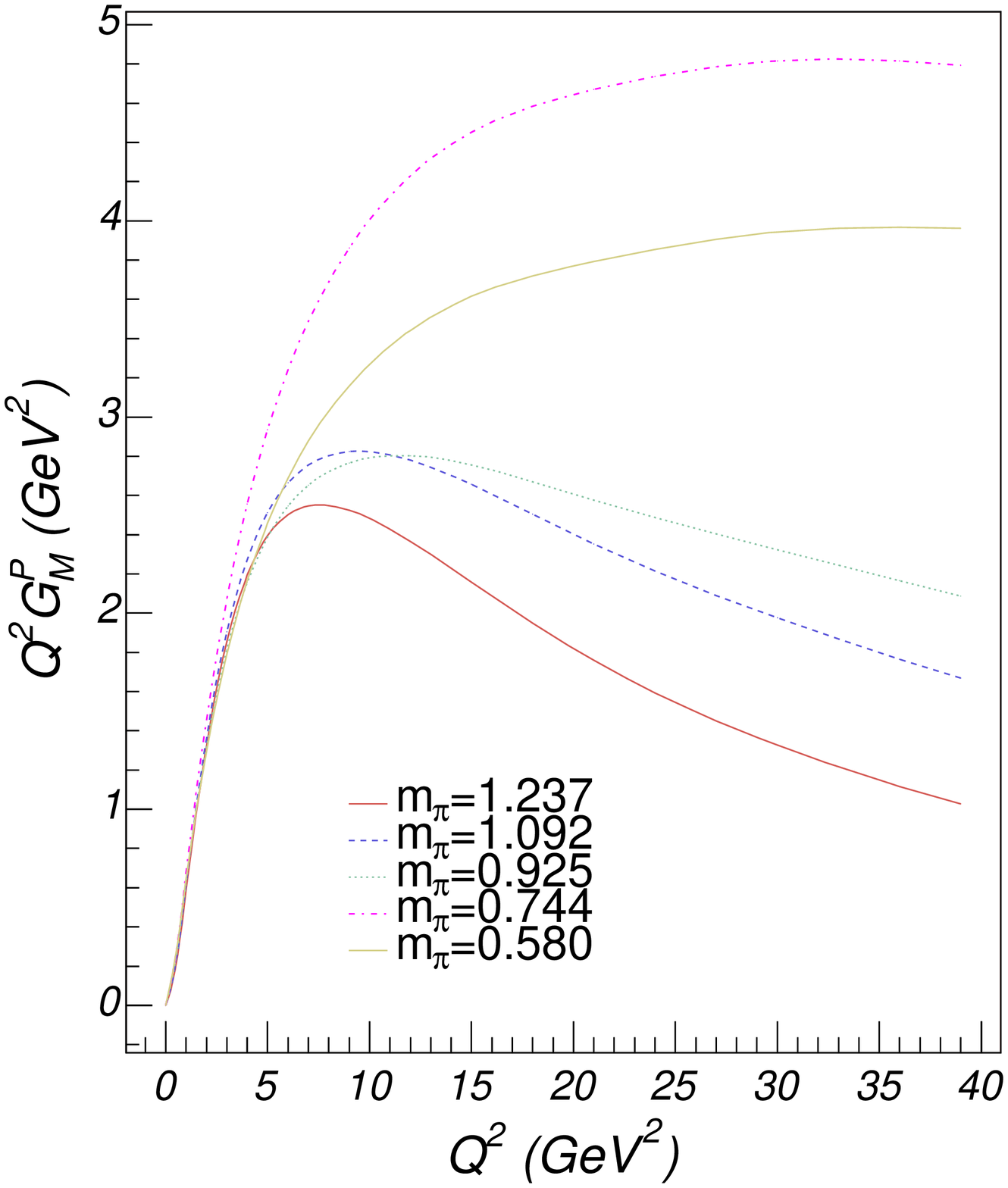}%
\caption{(Color online) LFCBM calculations using parameters (Figs.
\ref{PLOT_GAMMA_VS_MPI_6_0}, \ref{PLOT_GAMMA_VS_MPI_6_4} and
\ref{PLOT_MCHI_VS_LAT_A}) obtained by fitting the lattice results for the
proton magnetic form factor, $G_{M}$, at lattice \ spacing $a=0.26~GeV^{-1}$.}%
\label{PLOT_COMPARE_LAT_PROT_GM_6_4}%
\end{center}
\end{figure}

\begin{figure}
[ptbptbptb]
\begin{center}
\includegraphics[
height=7.8395in,
width=6.3235in
]%
{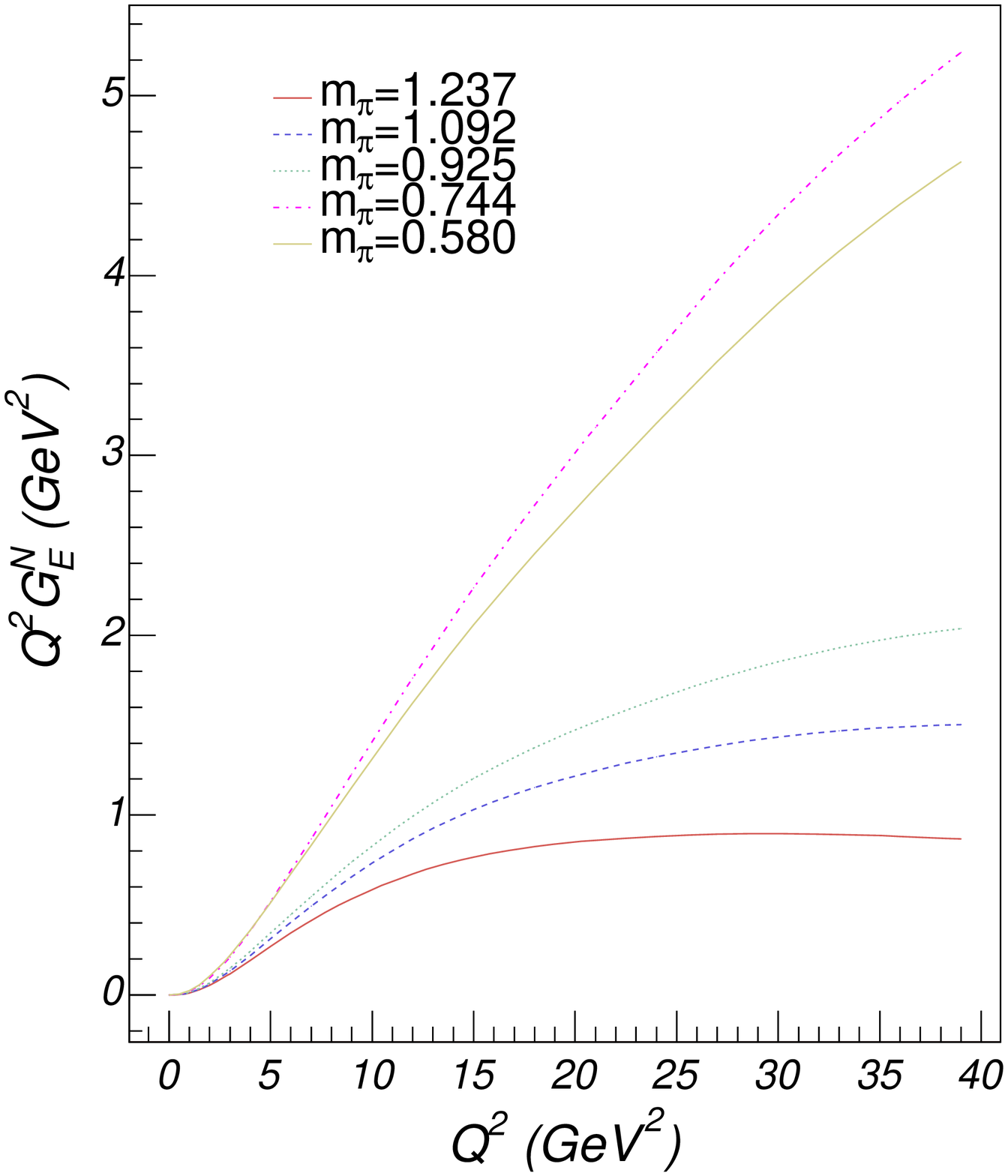}%
\caption{(Color online) LFCBM calculations using parameters
(Figs.\ref{PLOT_GAMMA_VS_MPI_6_0}, \ref{PLOT_GAMMA_VS_MPI_6_4} and
\ref{PLOT_MCHI_VS_LAT_A}) obtained by fitting the lattice results for the
neutron electric form factor, $G_{E}$, at lattice \ spacing $a=0.26~GeV^{-1}%
$.}%
\label{PLOT_COMPARE_LAT_NEUT_GE_6_4}%
\end{center}
\end{figure}

\begin{figure}
[ptbptbptbptb]
\begin{center}
\includegraphics[
height=7.7608in,
width=6.3235in
]%
{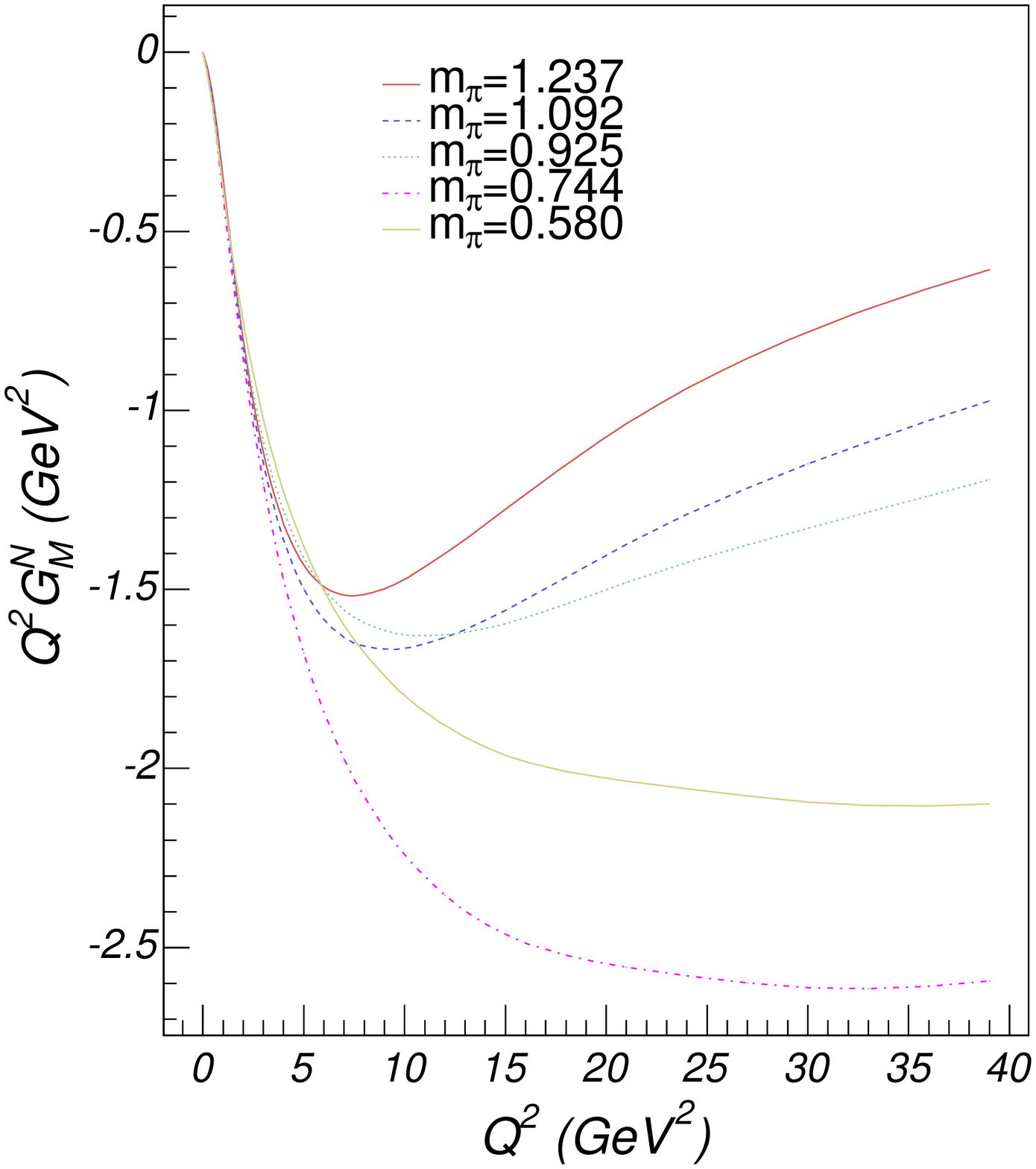}%
\caption{(Color online) LFCBM calculations using parameters
(Figs.\ref{PLOT_GAMMA_VS_MPI_6_0}, \ref{PLOT_GAMMA_VS_MPI_6_4} and
\ref{PLOT_MCHI_VS_LAT_A}) obtained by fitting the lattice results for the
neutron magnetic form factor, $G_{M}$, at lattice \ spacing $a=0.26~GeV^{-1}%
$.}%
\label{PLOT_COMPARE_LAT_NEUT_GM_6_4}%
\end{center}
\end{figure}

\begin{figure}
[ptbptbptbptbptb]
\begin{center}
\includegraphics[
height=7.8395in,
width=6.3235in
]%
{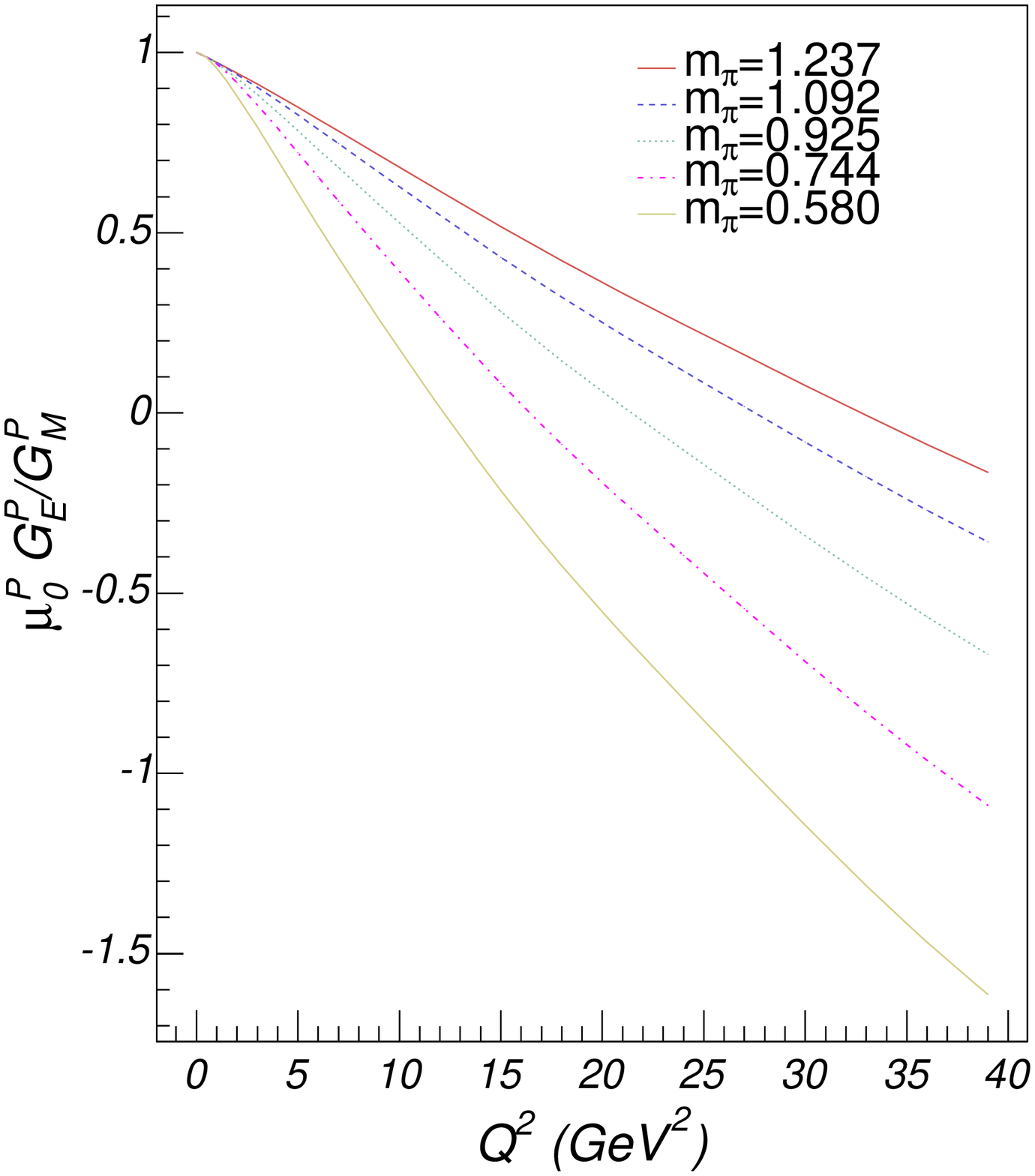}%
\caption{(Color online) LFCBM calculations using parameters
(Figs.\ref{PLOT_GAMMA_VS_MPI_6_0}, \ref{PLOT_GAMMA_VS_MPI_6_4} and
\ref{PLOT_MCHI_VS_LAT_A}) obtained by reproducing lattice results for the
ratio of proton form factors, $\mu_{0}G_{E}/G_{M}$, at lattice \ spacing
$a=0.26~GeV^{-1}$. }%
\label{PLOT_COMPARE_LAT_PROT_6_4}%
\end{center}
\end{figure}

\begin{figure}
[ptbptbptbptbptbptb]
\begin{center}
\includegraphics[
trim=0.000000in 0.000000in 0.002361in 0.000000in,
height=7.8395in,
width=6.3209in
]%
{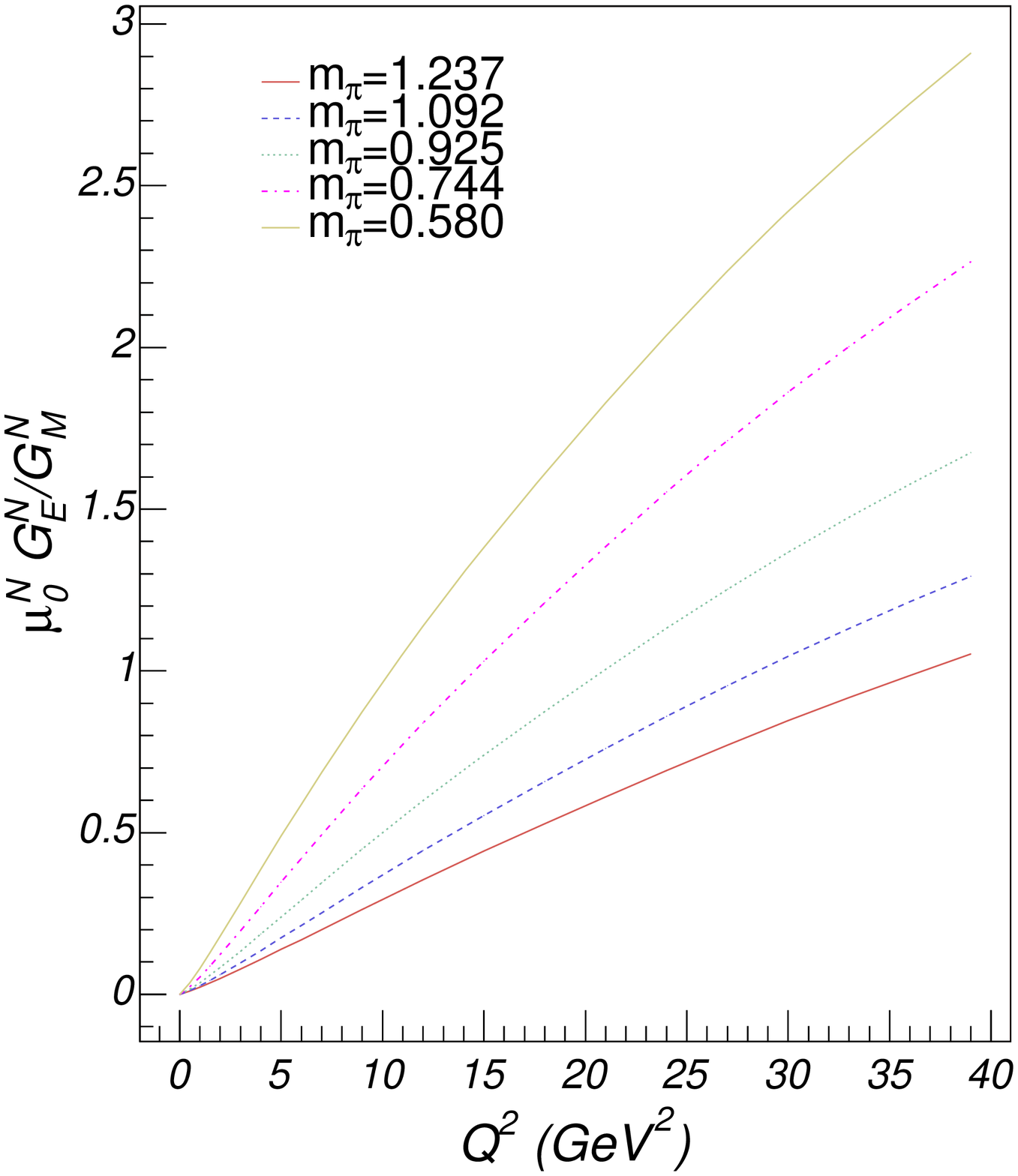}%
\caption{(Color online) LFCBM calculations using parameters
(Figs.\ref{PLOT_GAMMA_VS_MPI_6_0}, \ref{PLOT_GAMMA_VS_MPI_6_4} and
\ref{PLOT_MCHI_VS_LAT_A}) obtained by reproducing lattice results for the
ratio of neutron form factors, $\mu_{0}G_{E}/G_{M}$, at lattice \ spacing
$a=0.26~GeV^{-1}$. }%
\label{PLOT_COMPARE_LAT_NEUT_6_4}%
\end{center}
\end{figure}

Use of the values of $\gamma,M$ determined by the lattice data in the LFCBM
defines a lattice version of the LFCBM. We may use this new model to compute
the form factors at arbitrarily large values of $Q^{2}$, thereby extending the
kinematic range of the lattice calculations. The results are shown in
Figs.~\ref{PLOT_COMPARE_LAT_PROT_GE_6_4}, \ref{PLOT_COMPARE_LAT_PROT_GM_6_4},
\ref{PLOT_COMPARE_LAT_NEUT_GE_6_4} and \ref{PLOT_COMPARE_LAT_NEUT_GM_6_4}. In
Figs. \ref{PLOT_COMPARE_LAT_PROT_6_4} and \ref{PLOT_COMPARE_LAT_NEUT_6_4} we
show the corresponding plots of $\mu_{0}G_{E}/G_{M}$.

\subsection{Results at the Physical Pion Mass and Comparison With
Experiment\label{M_PHYS}}

\begin{figure}
[ptb]
\begin{center}
\includegraphics[
height=7.8118in,
width=6.2958in
]%
{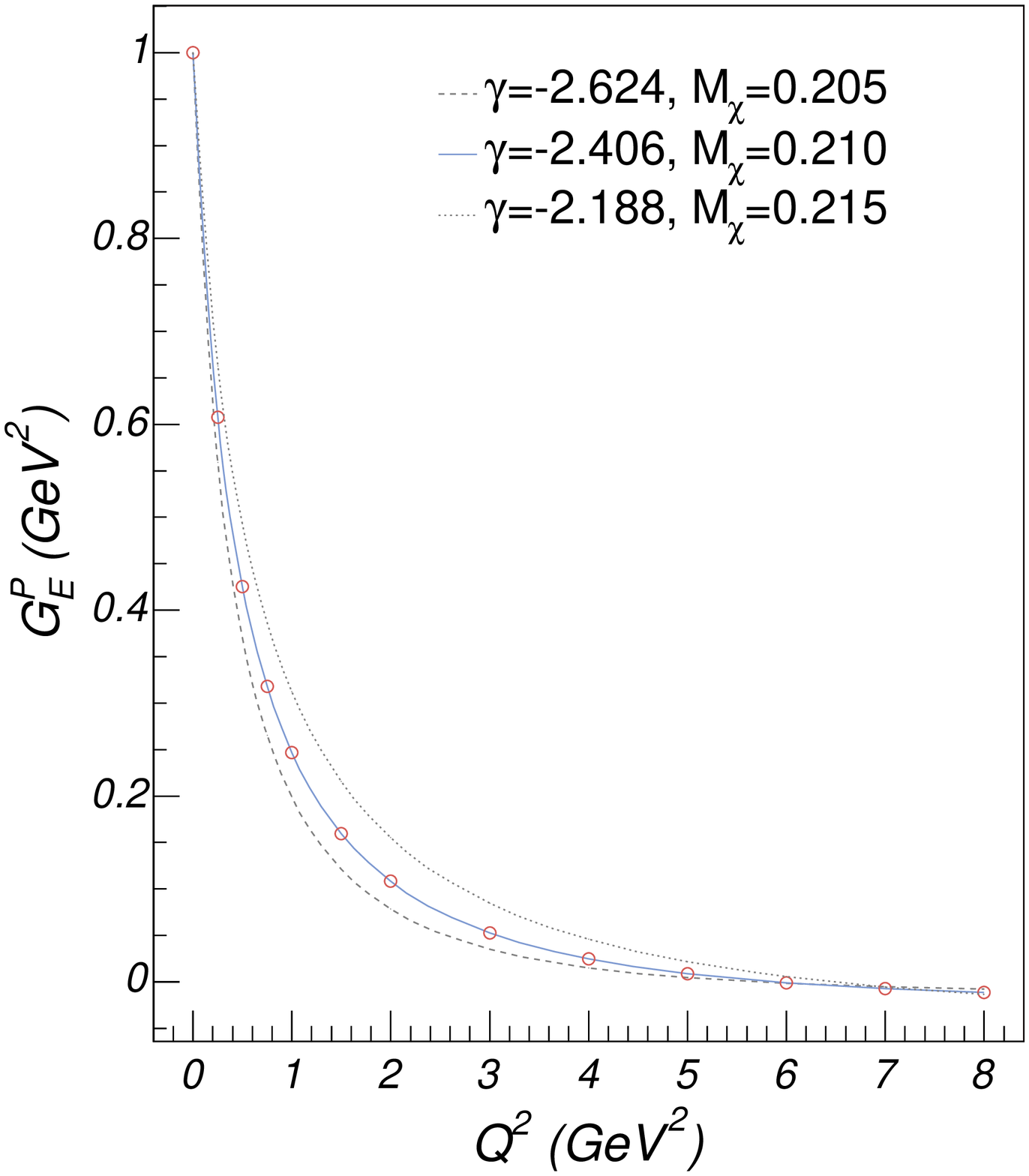}%
\caption{(Color online) Extrapolated calculations for the proton electric form
factor, $G_{E}$, for lattice spacing $a=0.26~GeV^{-1}$. The dashed and dotted
curves show the upper and lower limits of variation of the calculated values
due to the uncertainties of the parameters $\gamma$ and $M_{\chi}$.}%
\label{PLOT_PHYS_GE_PROT_6_4}%
\end{center}
\end{figure}

\begin{figure}
[ptbptbptb]
\begin{center}
\includegraphics[
height=7.8395in,
width=6.3235in
]%
{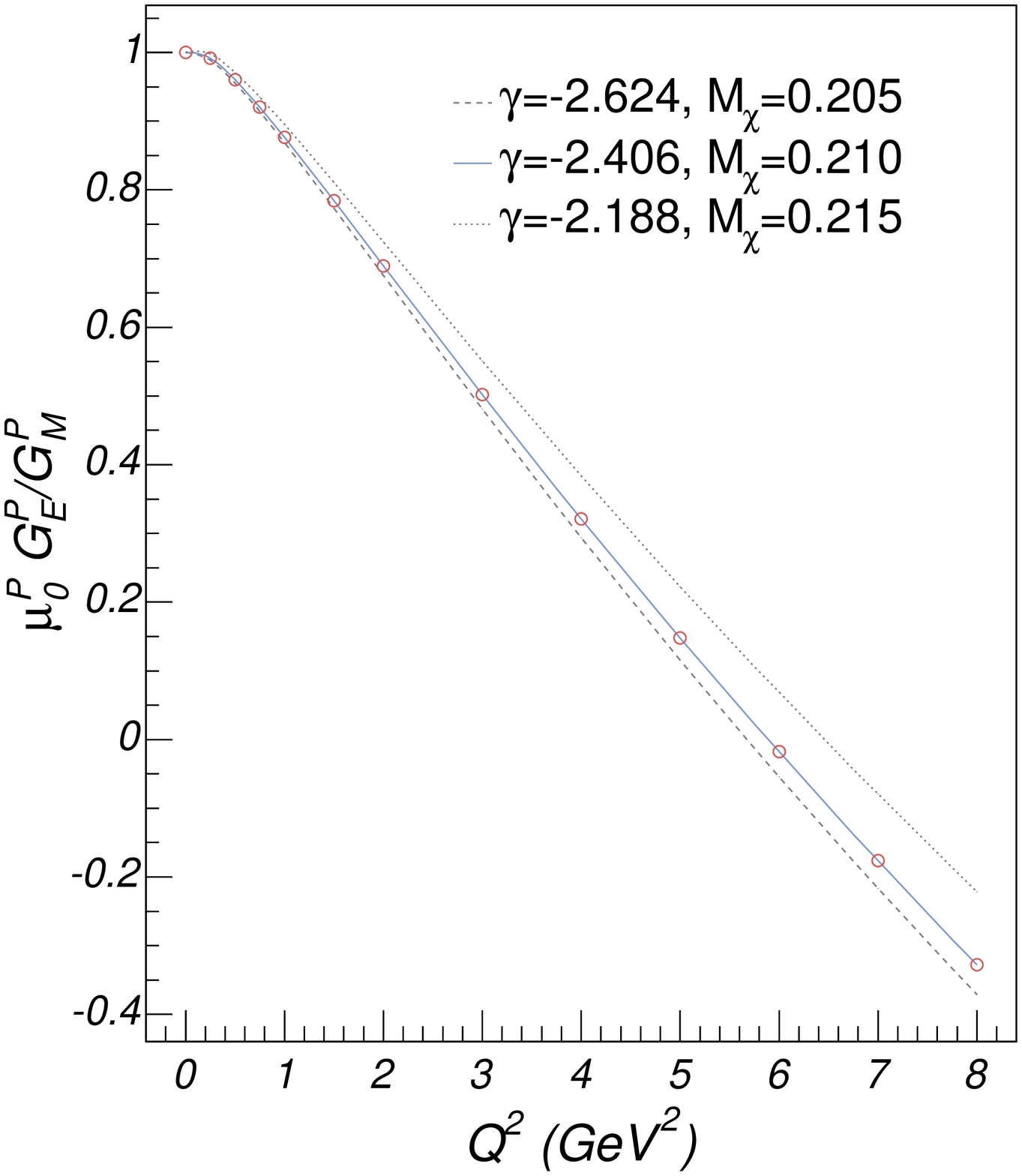}%
\caption{(Color online) Extrapolated calculations for the ratio of proton form
factors, $\mu_{0}G_{E}/G_{M}$, for lattice spacing $a=0.26~GeV^{-1}$. The
dashed and dotted curves show the upper and lower limits of variation of the
calculated values due to the uncertainties of the parameters $\gamma$ and
$M_{\chi}$.}%
\label{PLOT_PHYS_GE/GM_PROT_6_4}%
\end{center}
\end{figure}

\begin{figure}
[ptbptbptbptbptbptb]
\begin{center}
\includegraphics[
height=7.8395in,
width=6.3235in
]%
{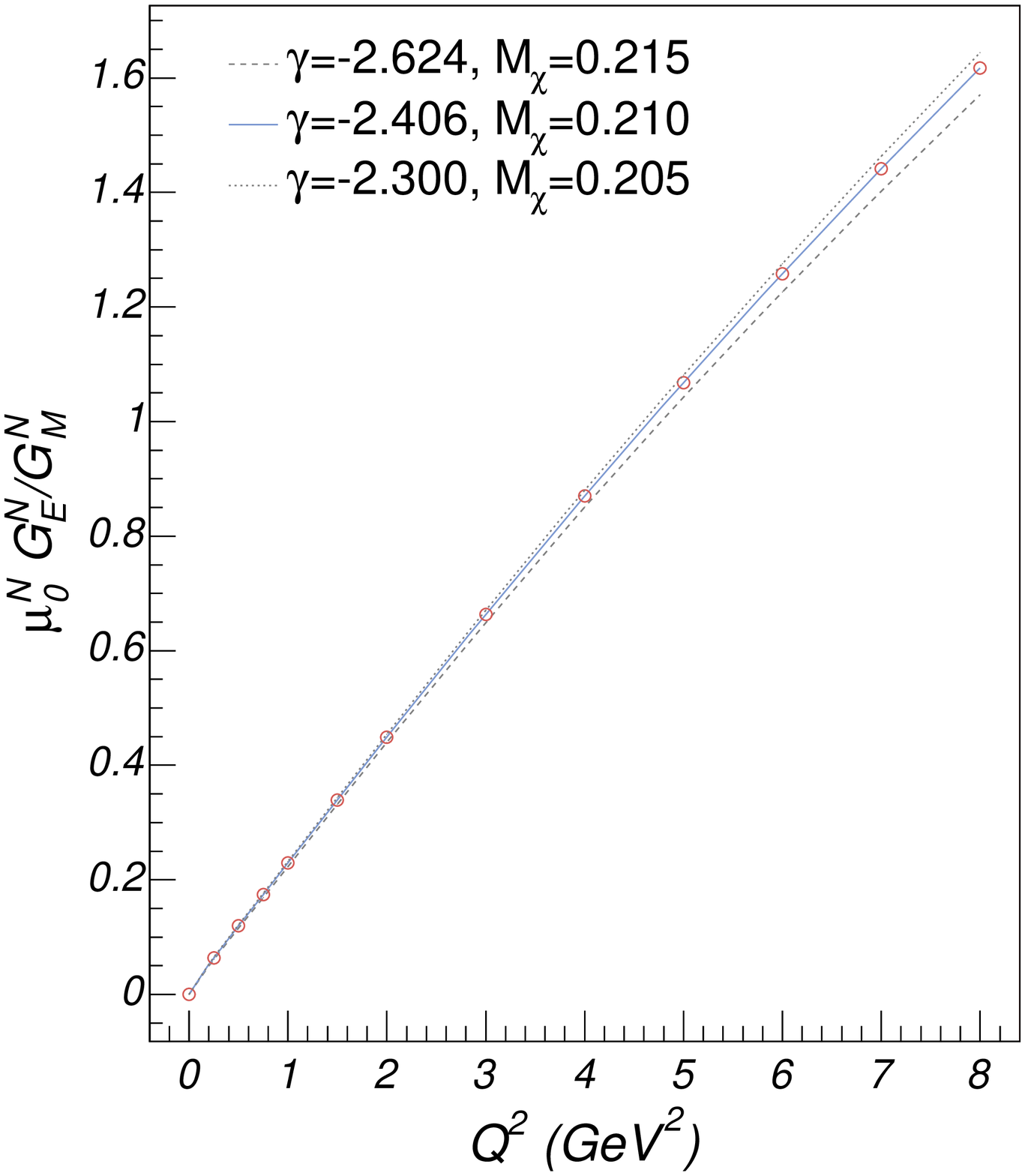}%
\caption{(Color online) Extrapolated calculations for the ratio of neutron
form factors, $\mu_{0}G_{E}/G_{M}$, for lattice spacing $a=0.26~GeV^{-1}$. The
dashed and dotted curves show the upper and lower limits of variation of the
calculated values due to the uncertainties of the parameters $\gamma$ and
$M_{\chi}$.}%
\label{PLOT_PHYS_GE/GM_NEUT_6_4}%
\end{center}
\end{figure}

We use the extrapolated values of $\gamma$ and $M$
(Figs.~\ref{PLOT_GAMMA_VS_MPI_6_0}-\ref{PLOT_MCHI_VS_LAT_A}) to calculate the
nucleon electric and magnetic form factors using the physical pion and nucleon
masses. The resulting plots for $G_{E}$, $G_{M}$ and their ratios vs. $Q^{2}$
for both proton and neutron are shown in Figs.~\ref{PLOT_PHYS_GE_PROT_6_4}%
-\ref{PLOT_PHYS_GE/GM_NEUT_6_4}. Figure~\ref{PLOT_PHYS_GE/GM_PROT_6_4} shows
that our results are in\ more or less good agreement with the experimental
data in the low-$Q^{2}$ region, but yield a slightly lower value of $Q^{2}$
for the zero cross-over point than that extrapolated from
experiment~\cite{GE/GM_0_MELNICHUK}. A new analysis that includes an estimate
of all of the effects of two photon exchange yields a zero-crossing value that
is somewhat closer to ours~\cite{GE/GM-0-CROSS} but future data will resolve
this unambiguously.
\begin{figure}
[ptbptb]
\begin{center}
\includegraphics[
height=7.8118in,
width=6.2958in
]%
{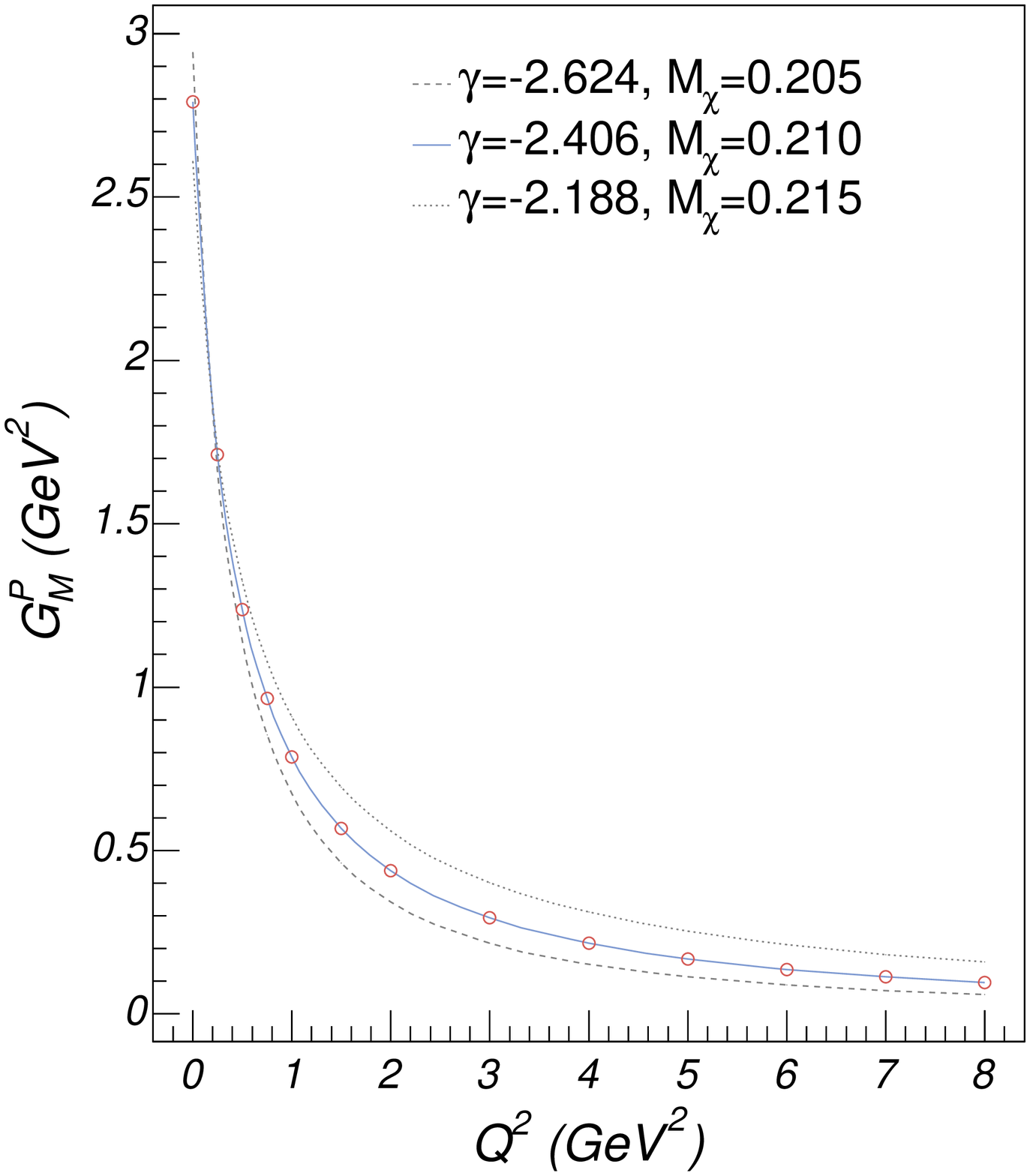}%
\caption{(Color online) Extrapolated calculations for the proton magnetic form
factor, $G_{M}$, for lattice spacing $a=0.26~GeV^{-1}$. The dashed and dotted
curves show the upper and lower limits of variation of the calculated values
due to the uncertainties of the parameters $\gamma$ and $M_{\chi}$.}%
\label{PLOT_PHYS_GM_PROT_6_4}%
\end{center}
\end{figure}
\begin{figure}
[ptbptbptbptb]
\begin{center}
\includegraphics[
height=7.7617in,
width=6.3235in
]%
{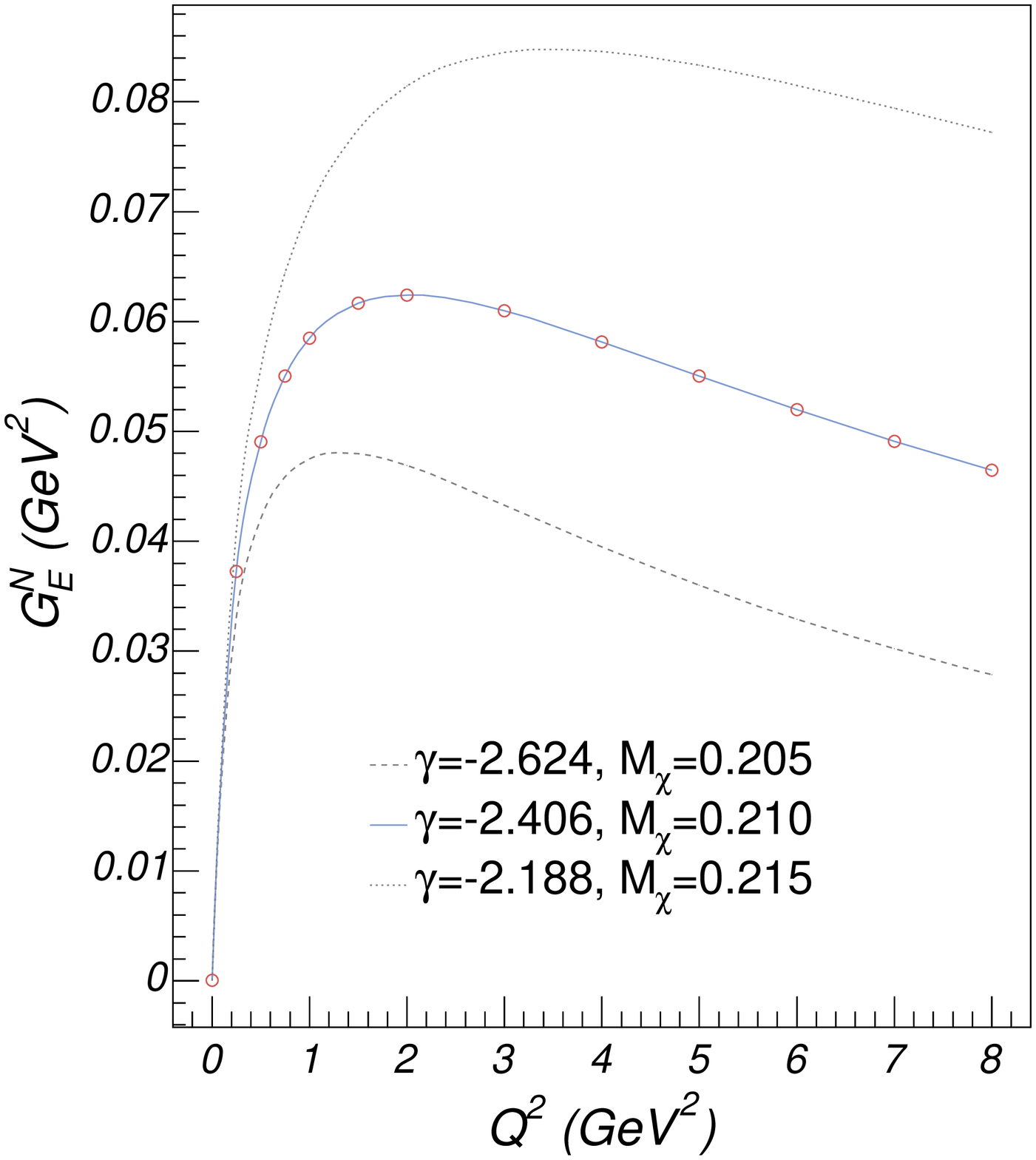}%
\caption{(Color online) Extrapolated calculations for the neutron electric
form factor, $G_{E}$, for lattice spacing $a=0.26~GeV^{-1}$. The dashed and
dotted curves show the upper and lower limits of variation of the calculated
values due to the uncertainties of the parameters $\gamma$ and $M_{\chi}$.}%
\label{PLOT_PHYS_GE_NEUT_6_4}%
\end{center}
\end{figure}
\begin{figure}
[ptbptbptbptbptb]
\begin{center}
\includegraphics[
height=7.7323in,
width=6.2958in
]%
{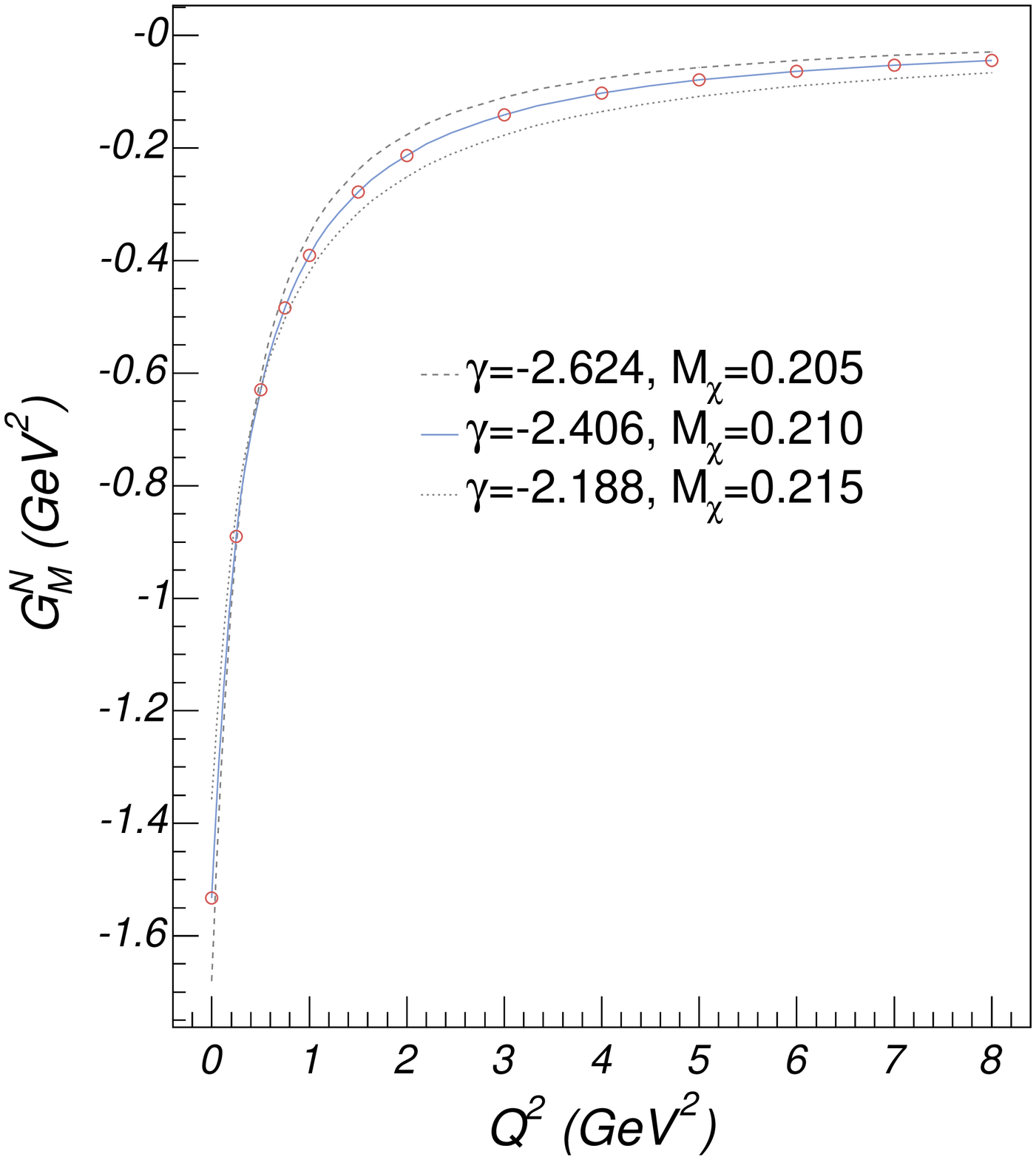}%
\caption{(Color online) Extrapolated calculations for the neutron magnetic
form factor,$\ G_{M}$, for lattice spacing $a=0.26~GeV^{-1}$. The dashed and
dotted curves show the upper and lower limits of variation of the calculated
values due to the uncertainties of the parameters $\gamma$ and $M_{\chi}$.}%
\label{PLOT_PHYS_GM_NEUT_6_4}%
\end{center}
\end{figure}

\begin{figure}
[ptb]
\begin{center}
\includegraphics[
height=4.8144in,
width=5.0566in
]%
{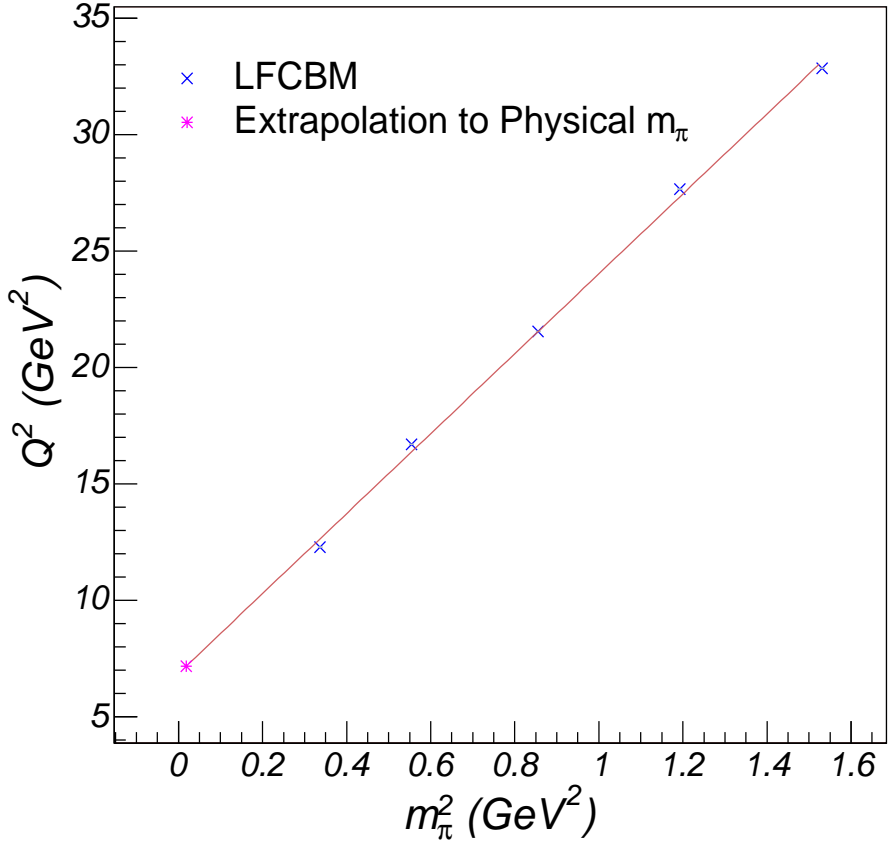}%
\caption{(Color online) Extrapolation of $Q_{Cross}^{2}$ where $G_{E}/G_{M}$
passes through zero for lattice spacing $a=0.26~GeV^{-1}$.\ }%
\label{PLOT_GE_GM_0_INTERCEPT}%
\end{center}
\end{figure}

An alternative method of determining the value of the $Q^{2}$ for which
$G_{E}/G_{M}$ passes through zero at the physical pion mass is to fit the
crossover values as a linear function of $m_{\pi}^{2}$ and extrapolate again
to the physical pion mass. The resulting plot is shown in {}%
Fig.~\ref{PLOT_GE_GM_0_INTERCEPT}.%
This procedure yields approximately the same cross-over point as found in
Fig.~\ref{PLOT_PHYS_GE/GM_PROT_6_4}.

\section{Discussion\label{RESULTS_AND_MODEL}}

Our study of the form factors calculated using the LFCBM with parameters
determined by lattice data and by extrapolation to the physical pion masses
yields very interesting results. The ratio $G_{E}^{p}/G_{M}^{p}$ passes
through zero for all of the calculations. The main variation of the position
of the crossover between the fitting curves shown in
Figs.~\ref{PLOT_COMPARE_LAT_PROT_6_4} and \ref{PLOT_COMPARE_LAT_NEUT_6_4}
comes from the variation of the nucleon mass, and not the variation of
$\gamma$. Even though for the physical pion mass, the ratio varies rapidly as
a function of $\gamma$ in the region $\gamma\thicksim-2$, the function
$G_{E}/G_{M}$ for the neutron has a turning point at about $\gamma
\thicksim-2.3$. We shall explain these features using the LFCBM.

Let us express the ratio $G_{E}/G_{M}$ in terms of Pauli and Dirac form
factors, $F_{1}$ and $F_{2}$, respectively, using Eq.~(\ref{FORMULA_G_TO_F})%
\begin{equation}
\frac{G_{E}}{G_{M}}=\frac{F_{1}-Q^{2}/(4M_{N}^{2})F_{2}}{F_{1}+F_{2}}%
=1-\frac{1+Q^{2}/(4M_{N}^{2})}{1+F_{1}/F_{2}} \label{FORMULA_GE/GM_EXPAND}%
\end{equation}
Consider first the values of $Q^{2}=(Q^{2}_{\mathrm{Cross}})$ where the ratio
$G_{E}/G_{M}$ for the proton passes through zero for the set of calculations
shown in Fig. \ref{PLOT_COMPARE_LAT_PROT_6_4}.
Equation~(\ref{FORMULA_GE/GM_EXPAND}) tells us that
\begin{equation}
Q_{\mathrm{Cross}}^{2}=4M_{N}^{2}\frac{F_{1}}{F_{2}}
\label{FORMULA_Q2_CROSS_OVER}%
\end{equation}

Now let us consider the formula for $F_{i\alpha}(Q^{2})$,
Eq.~(\ref{FORMULA_F_TO_F0_FB_FC}). The second and third terms in
Eq.~(\ref{FORMULA_F_TO_F0_FB_FC}) are only significant in the $low-Q^{2}$
region for physical pion masses. In the $high-Q^{2}$ region, or for lattice
calculations with high pion mass, these terms are vanishingly small. Indeed
the numerical calculations support these statements, so we can neglect their
contribution in the rest of the discussion.

The corresponding formulas for $F_{1}^{(0)}$ and $F_{2}^{(0)}$ from
Ref.~\cite{Miller:2002qb} are%
\begin{equation}
F_{1}^{(0)}(Q^{2})=\int\frac{d^{2}q_{\bot}d\xi}{\xi(1-\xi)}\frac{d^{2}K_{\bot
}d\eta}{\eta(1-\eta)}\tilde{\Phi}^{\dagger}(M_{0}^{\prime})\tilde{\Phi}%
(M_{0})\times\left\langle \chi_{0}^{rel}(\mathbf{p}_{1}^{\prime}%
,\mathbf{p}_{2}^{\prime})|\chi_{0}^{rel}(\mathbf{p}_{1},\mathbf{p}%
_{2})\right\rangle \left\langle \uparrow\mathbf{p}_{3}^{\prime}|\uparrow
\mathbf{p}_{3}\right\rangle \label{FORMULA_F10_PROT}%
\end{equation}%
\begin{equation}
{\frac{QF_{2}^{(0)}(Q^{2})}{2M_{N}}}=\int\frac{d^{2}q_{\bot}d\xi}{\xi(1-\xi
)}\frac{d^{2}K_{\bot}d\eta}{\eta(1-\eta)}\tilde{\Phi}^{\dagger}(M_{0}^{\prime
})\tilde{\Phi}(M_{0})\times\left\langle \chi_{0}^{rel}(\mathbf{p}_{1}^{\prime
},\mathbf{p}_{2}^{\prime})|\chi_{0}^{rel}(\mathbf{p}_{1},\mathbf{p}%
_{2})\right\rangle \left\langle \uparrow\mathbf{p}_{3}^{\prime}|\downarrow
\mathbf{p}_{3}\right\rangle \label{FORMULA_F20_PROT}%
\end{equation}
The $\tilde{\Phi}(M_{0})$ factors are wave functions of the form of
Eq.~(\ref{FORMULA_SPAT_WAV_FUNC}), but using the lattice values of $\gamma,M$
shown in Figs.\ref{PLOT_GAMMA_VS_MPI_6_0}-\ref{PLOT_MCHI_VS_LAT_A}. We stress
that these two integrals differ only by the last factor, which gives the spin
non-flip and spin-flip dependence of $F_{1}^{(0)}$ and $QF_{2}^{(0)}/2M_{N}$,
respectively. At high $Q^{2}$ these matrix elements are each of order $Q$,
causing the ratio $QF_{2}/F_{1}$ to be approximately constant. So we can
express $Q_{Cross}^{2}$ as%
\begin{equation}
Q_{Cross}^{2}=4M_{N}^{2}\left(  \frac{\int\nolimits_{\uparrow\uparrow}}%
{\int_{\uparrow\downarrow}}\right)  ^{2} \label{FORMULA_Q2_CROSS_VIA_INTGL}%
\end{equation}
where $\int\nolimits_{\uparrow\uparrow}$ denotes the integral for $F_{1}%
^{(0)}$, and $\int_{\uparrow\downarrow}$ denotes the integral for $F_{2}%
^{(0)}$.

\begin{figure}
[ptb]
\begin{center}
\includegraphics[
height=4.5186in,
width=4.7547in
]%
{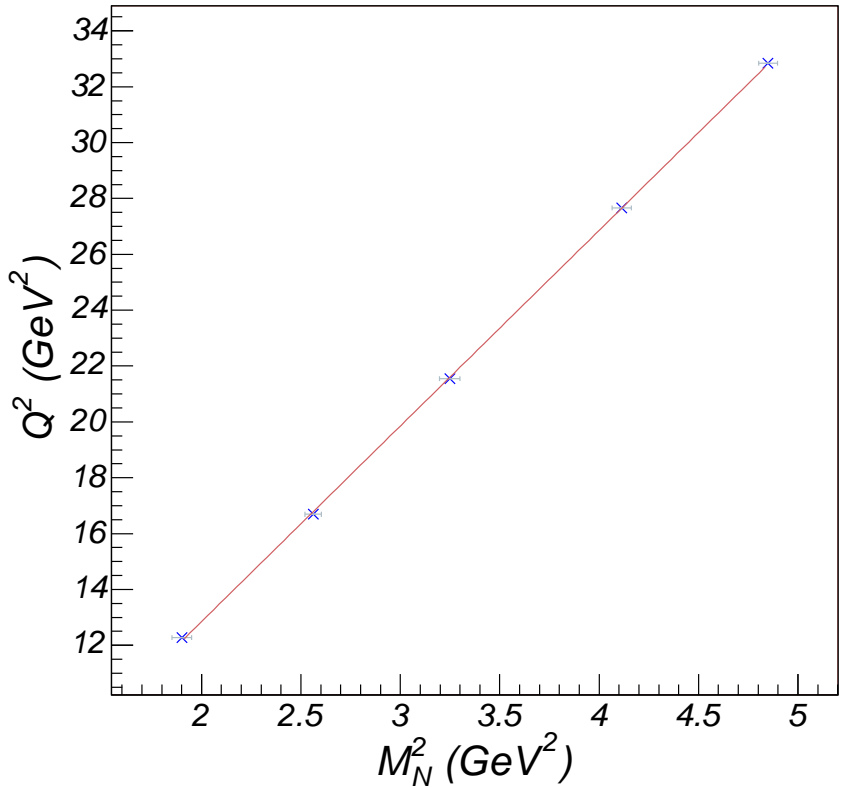}%
\caption{(Color online) Linear fit of $Q_{Cross}^{2}$ where $G_{E}/G_{M}$
passes through zero for lattice spacing $a=0.26~GeV^{-1}$.\ }%
\label{PLOT_GE_GM_0_INTERCEPT_VS_MNUCL}%
\end{center}
\end{figure}

In the high-$Q^{2}$ region the ratio in Eq.~(\ref{FORMULA_Q2_CROSS_VIA_INTGL})
is approximately a constant, because the difference comes only from the
overlap factors of the spin-dependent parts of the wave functions in the
integrals (See \cite{Miller:2002ig},\cite{Miller:2003sa}). So the behavior of
$Q_{Cross}^{2}$ is governed primarily by the factor $M_{N}^{2}$. The linear
variation of $Q_{Cross}^{2}$ vs. $M_{N}^{2}$ presented in
Fig.~\ref{PLOT_GE_GM_0_INTERCEPT_VS_MNUCL} shows the validity of this
interpretation.

We can also understand the behavior of $G_{E}/G_{M}$ versus $\gamma$, by
considering its role in the wave function. The factor $\gamma$ determines the
size of the momenta appearing in the integrands of Eqs.(\ref{FORMULA_F10_PROT}%
) and (\ref{FORMULA_F20_PROT}). The corresponding integrands differ by terms
that are ratios of second order polynomials of the integration variables. For
large absolute values of $\gamma$, the high momenta are cut off more strongly,
so that the contribution of terms that cause differences between the integrals
are not very significant. For small absolute values of $\gamma$ the integrals
become more sensitive to those terms and we obtain a larger variation of the
ratios of the integrals and hence the ratio $G_{E}/G_{M}$.

\section{Conclusion}

We have seen that the LFCBM can produce a very good description of the lattice
QCD data for the nucleon form factors over a wide range of quark masses with a
smooth, analytic variation of the wave function parameter, $\gamma$, and the
constituent quark mass, $M$. The pion cloud plays very little role in the mass
range for which the lattice simulations have been made but it rapidly becomes
more important as we approach the chiral limit. From the rather strong
dependence of the form factors on the lattice spacing, $a$, it is not yet
clear that we have obtained a good approximation to the continuum limit, but
the form factors obtained at the smallest value of $a$ are in reasonable
agreement with experimental data in the low-$Q^{2}$ region for which the
lattice simulations were made.

At present the lattice simulations are limited to values of the momentum
transfer at or below $2%
\operatorname{GeV}%
^{2}$ and it is therefore a very big extrapolation to look at the behavior of
the form factors in the region of greatest current interest. Nevertheless, the
behavior of $G_{E}/G_{M}$ which we find is particularly interesting. The ratio
crosses zero for all values of the quark mass but the position where this
happens varies over a very wide range of $Q^{2}$. This variation can be
understood almost entirely in terms of the variation of the corresponding
nucleon mass, given that the ratio $QF_{1}/F_{2}$ is approximately $Q^{2}$
independent in the model. We obtain the same value of $Q^{2}$ for the
cross-over whether we extrapolate the position as a function of quark mass or
simply evaluate the form factors at the physical pion mass, using the fitted
dependence of the wave function parameters on pion mass.

In the immediate future it is clearly very important to improve on the lattice
data, both by ensuring that we really have a good approximation to the
continuum limit (e.g., by using a suitably improved action) and by extending
the calculations to higher values of $Q^{2}$. It would also be important to
remove the need for quenching, even though that may not be such a limitation
at large $Q^{2}$. From the point of view of developing a deeper understanding
of QCD itself it is important that the LFCBM is able to describe the present
lattice data over such a wide range of masses. We would encourage a similar
exercise for other models as a novel test of their validity. It remains to be
seen whether the LFCBM has indeed been successful in predicting the behavior
of the form factors at higher $Q^{2}$ and indeed whether it will match future
experimental data.

\section*{Acknowledgements}

This work was supported in part by the U.S. National Science Foundation (Grant
Number 0140300), the Southeastern Universities Research Association(SURA), DOE
grant DE-AC05-84ER40150, under which SURA operates Jefferson Lab, and also DOE
grant DE-FG02-97ER41014. G.~A.~M. thanks Jefferson Lab for its hospitality
during the course of this work. H. H. M. thanks the Graduate School of
Louisiana State University for a fellowship partially supporting his research.

\end{document}